\documentclass[aps,prb,showpacs,12pt]{revtex4-1}
\usepackage{graphicx}
\usepackage{xcolor}
\usepackage{bm,amsmath,amssymb}
\newcommand{\bos}{\boldsymbol}

\begin{document}
\title{On the thermal properties of knotted block copolymer rings }
\author{Neda Abbasi Taklimi$^1$}
\email{neda.abbasi\_taklimi@phd.usz.edu.pl}
\author{Franco Ferrari$^1$}
\email{franco@feynman.fiz.univ.szczecin.pl}
\author{Marcin Rados{\l}aw Pi\c{a}tek$^1$}
\email{marcin.piatek@usz.edu.pl}
\author{Luca Tubiana$^{2.3.4}$}
\email{luca.tubiana@unitn.it}
\affiliation{$^1$CASA* and Institute of Physics, University of Szczecin,
  Szczecin, Poland} 
\affiliation{$^2$ Physics Department, University of Trento, Via
  Sommarive 14, I-38123, Trento, Italy}
\affiliation{$^3$ INFN-TIFPA, Trento Institute for Fundamental Physics and Applications, I-38123 Trento, Italy}
\affiliation{$^4$ Faculty of Physics, University of Vienna, Boltzmanngasse 5, 1090 Vienna, Austria}
\date{\today}

\begin{abstract}
The thermal properties of coarse grained knotted polymers containing two
kinds of monomers $A$ and $B$ fluctuating in a solution are investigated on a simple cubic lattice using the Wang-Landau MC algorithm.
These knots have a more complex phase diagram than knots formed by homopolymers, including the possible presence of metastable states.
Two different setups are considered: i) charged block copolymers in a ion solution and ii) neutral copolymers with the $A$ monomers above and the $B$ monomers below the theta point.
A precise interpretation of the peaks observed in the plots of the specific heat capacity is provided.
In view of possible  applications  in medicine and the construction of
intelligent materials, it is also shown that the behavior of copolymer rings
can be tuned by changing both their monomer configuration and topology.
We find that the most stable compact states are formed by charged copolymers in which very short segments with $A$ monomers are alternated by short segments with $B$ monomers. In such knots the transition from the compact to the expanded state is very fast, leading to a narrow and high peak in the specific heat capacity which appears at very high temperatures. 
The effects of topology allow to tune the radius of gyration of the knotted polymer ring and to increase or decrease the temperatures at which the observed phase transitions or rearrangements of the system occur. While we observe a general fading out of the influence of topology in longer polymers, our simulations have captured a few exceptions to this rule.
\end{abstract}
\maketitle
%\clearpage
\section{Introduction}\label{introd}
Polymer knots are abundant in nature and in artificial polymer
materials \cite{sauvage,arsuaga,ramirez,amabilino,tezuka}. They can be created
in the laboratory \cite{amabilino,tezuka,leigh} and have attracted a 
considerable attention both from experimentalists and theoreticians of
several different disciplines including
chemistry\cite{chemistryarticles1,StauchDreuw,Schaufelberger},
engineering\cite{engineeringarticles1}, 
mathematics \cite{mathematicalarticles2,mathematicalarticles1}  and
physics\cite{physicalarticles1,physicalarticles2,physicalarticles3,physicalarticles4,janke3}. In this work
we consider the static properties of knots
made by copolymers. We study in particular diblock copolymers
consisting of polymers with two different kinds of monomers $A$ and
$B$.
Part of the  motivations
for this work come from biology. In fact, DNA and other biomolecules
are characterized by regions that have different 
properties and can thus be regarded as copolymers. Recently, a diblock
copolymer approximation of a piece of DNA has been used in order to
understand how the dishomogeneities in the flexibility affect
the localization of knots on a piece of circular DNA
\cite{orlandinibaiesizontaworkonstiffness,daietal}.  
The study of diblock copolymers can be helpful also in technological
applications. For instance, it is already known that the presence of
knots affects the behavior of  
polymer materials. 
Indeed, the
elasticity response of elastomers cannot be understood without
considering the fact that the polymer chains inside these materials
form knots and links.
The effects of the presence of knots in the conformational properties of ring $AB$ diblock-copolymers have  already been noted for instance in Ref.~\cite{vlahosetal}.

\begin{figure}
  \begin{center}
    \includegraphics[height=5cm]{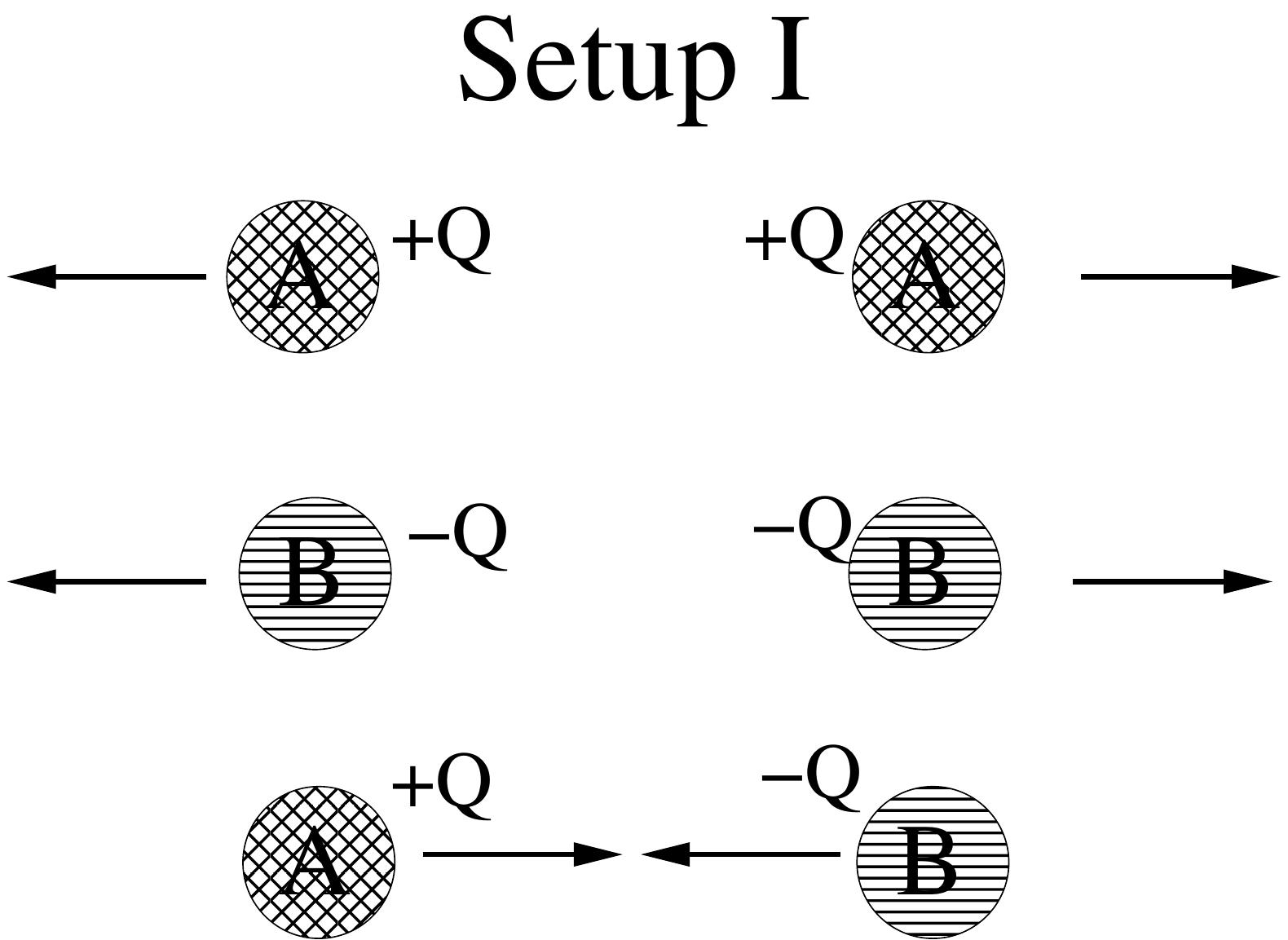}\hspace*{3cm}
    \includegraphics[height=5cm]{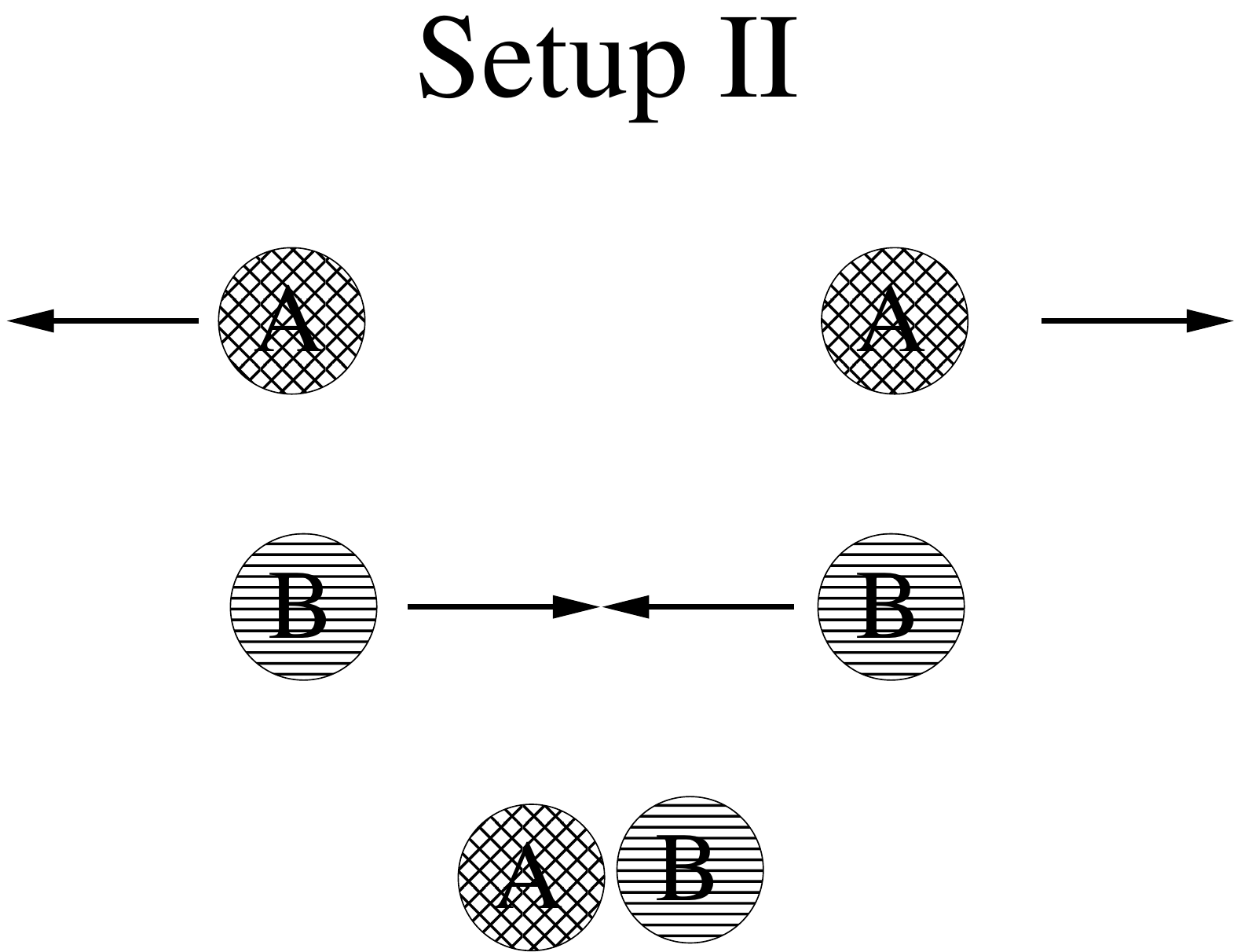}
    \caption{This figure summarises the main features of the Setups I and II considered in this paper. Particles in Setup I have a charge $Q$. Apart from a proportionality factor, we have that $Q\propto\sqrt{\varepsilon}$. The exact value of the proportionality constant is not relevant for performing the calculations. Arrows explain if the interactions are repulsive or attractive.
      In Setup II the $A$ and $B$ monomers are subjected only to excluded-volume interactions, so that arrows are not necessary in this case.
 }\label{scheme-setups} 
  \end{center}
\end{figure}

The statistical mechanics of open or circular diblock
copolymers has been thouroghly investigated in the past, see e.~g.
\cite{marko,vilgis,holystvilgis,huber,metzler}. Polyelectrolytes similar to those treated here in Setup~I (see below) have been considered in \cite{velyaminov}. Setup~II has some similarities with the Hydrophilic-Polar (HP) protein model \cite{dill}, however in our case the $B$ monomers are subjected to attractive forces.  The HP model has been studied using the Wang-Landau algorithm in \cite{wuest,wuest2}.
More recently, there has been some interest on circular
diblock copolymers with non-trivial topologies
\cite{orlandinibaiesizontaworkonstiffness,daietal,vlahosetal,najafi,
  kuriatasikorski,benahmed,tagliabueetal,kumar}. For example, in \cite{tagliabueetal} it has been
investigated how the stiffness heterogeneity or the presence of charges
influence the localization of the knot. The role of stiffness and heterogeneity in knot production has been explored in Ref.~\cite{cardellietal}. Other aspects of the topology of diblock-copolymers have been treated in \cite{kumar}.
With the help of the Wang-Landau algorithm \cite{wl}, the statistical
mechanics of knotted diblock copolymers has been studied in
Refs.~\cite{wang,swetnam,FFBlockcopolymerknots-preprint}.  
The goal of the present work is to extend the results of \cite{wang,FFBlockcopolymerknots-preprint}
to longer polymers, showing that remarkable properties emerge in this
case.
Knotted copolymers
are defined here on a simple cubic lattice. Their monomers are
subjected to different kinds of  very short-range interactions
reproducing different physical setups. In one setup, called hereafter
Setup I, a charged polymer is fluctuating in an ion solution that
screens the long-range Coulomb interactions. 
Monomers of type $A$ have
a positive charge, while monomers of type $B$ are negatively charged. 
The upshot is that 
monomers of the same kind repel themselves, while the interactions
between the $A$ and $B$ monomers are attractive.
Setup I is also relevant for 
 the case of polymers in water. In water at room temperature, in
fact, the Bjerrum length $l_B$ amounts to just $7\AA$. Let us recall
that the constant $l_B$
measures
the length scale at which the strength of the Coulomb
interactions in a dielectric medium becomes equal to the thermal
energy $k_BT$, where $k_B$ is the Boltzmann constant and $T$ is the
temperature \cite{Dobrynin}.
In the second studied setup, that will be  named Setup II, the solvent
is good for
the monomers of type $A$, which thus repel themselves, while 
monomers of type $B$ are below the theta point and attract themselves.  
Between monomers of type $A$ and $B$ we suppose that only excluded
volume forces are acting.
Fig.~\ref{scheme-setups} summarises the main features of both setups.
Other setups are possible, see for
instance \cite{vilgis}.
%With this set-up it is possible to rule out  immediately  block
%copolymers  in the 
%configuration  $(AB)_n$, i.~e. 
%in which the block $AB$ is repeated $n$ times.
%In fact, on a simple cubic lattice the
%$A-$ and $B$ monomers would always interact with themselves, leaving
%no place for the attractive interactions between $A$ and $B$.
%With this set-up it is possible to rule out  immediately  block
%copolymers  in the 
%configuration  $(AB)_n$, i.~e. 
%in which the block $AB$ is repeated $n$ times.
%In fact, on a simple cubic lattice the
%$A-$ and $B$ monomers would always interact with themselves, leaving
%no place for the attractive interactions between $A$ and $B$.

Multiblock copolymers with different monomer distributions and
interactions are considered. 
We construct knots containing alternating units with $n_A$ monomers of type $A$
and $n_B$ monomers of type $B$ until the total
number of monomers $N$ is obtained. If $N$ is not a multiple of
$n_A+n_B$, a slight excess of monomers of type $A$ is allowed.  
In the following, it will be convenient to introduce the total
number $N_A$ of $A-$monomers and the total number $N_B$ of
$B-$monomers. Of course 
$N_A+N_B=N$. 
Multiblock copolymers of this kind will be denoted with the symbols
$M_I(N,n_A,n_B)$ and $M_{II}(N,n_A,n_B)$.
The subscripts  $I$ and $II$ refer the two setups discussed before.
A particularly interesting subcase is that
of
the $AB-$diblock 
copolymers composed by two segments, one with $A$ monomers and the other with $B$ monomers. $AB-$diblock copolymers will be distinguished within the more general class of multiblock copolymers introducing the new symbols
$D_I(N_A,N_B)$ and $D_{II}(N_A,N_B)$. Of course, $D_{I/II}(N_A,N_B)=M_{I/II}(N,N_A,N_B)$, where $I/II$ means Setup I or II.
Following the above conventions, uncharged homopolymer knots with
$N$ monomers can be listed under Setup II. A homopolymer knot in a good
solvent  corresponds to the $AB-$diblock copolymer $D_{II}(N,0)$ with zero monomers of type $B$,
while a  homopolymer knot  in a bad solvent can be described with the
symbol $D_{II}(0,N)$.
To further characterize the analyzed knots,  also their
monomer composition
$f=N_A/(N_A+N_B)$ will be used.

The rationale for investigating knots made by copolymers is to obtain
macromolecules with different properties by changing the knot
topology, the $A/B$ monomer ratio and the monomer distribution along
the chain. 
We  show here that this goal can indeed be achieved and that copolymer
knots exhibit 
a variety of behaviors that are absent in knots formed by homopolymers.
In a nutshell, it turns out that the relevant parameters of a polymer ring, like gyration radius, heights and temperatures of the peaks of the heat capacity, specific energy and number of contacts (non-contiguous monomers that are at the distance of one lattice unit, see below for a more precise definition), are highly affected by the monomer distribution.
Topology has strong effects on the behaviour of short polymers. In the case of longer polymers, these effects  fade out, but still the properties of the knot may be tuned by choosing knots with the same length and monomer distribution, but  different topology.
Despite the vanishing influence of topology with increasing polymer lengths, we have observed that, in particular cases, the thermal behaviour may drastically change depending on the type of the knot even in longer polymers.

Concluding this Introduction, we would like to stress that the present
analysis requires the sampling of knot conformations that are
very compact, so that the density of monomers is very high. These
compact states often include conformations that are extremely rare
and thus very difficult to be sampled using Monte Carlo
algorithms. Such conformations may act as bottlenecks in a Monte Carlo
simulation, relevantly increasing  the computation time. The problem
of handling rare events is not only related to the case of polymer
systems. In the Wang-Landau Monte Carlo algorithm, which we will adopt
for our simulations, this issue has been already treated in several
previous publications, see
e. g. \cite{rarestatesprevioispublications,dellago}.
Following \cite{publicationoflandauwithasimilaraccelerationtechnique},
in this work we have used parallelization techniques to speed up the Wang-Landau algorithm
allowing to study the lowest energy
conformations. The latter are important because they are dominating at extremely low temperatures. Some
of these techniques 
have been discussed in more details in
\cite{YZFFPhysicaA2017}.
In large systems the
number of states to be sampled is enormous. In a knot with $N=500$, for
instance, a conformation with the lowest observed energy value can
appear once  in a set of $10^{11}$ samples. In the case of a knot with
$N=1000$, the Wang-Landau sampling process requires a few months.
%For this reason, for longer polymers energy cut-offs have been
%introduced.
%As a strategy to be sure that the observed results are not depending
%on the choice of the cut-offs, several runs have been performed for
%the same knot with increasingly extended energy intervals. It has been checked %that the
%plots of the studied observables remain stable for sufficiently large energy
%intervals.

With the inclusion in our calculations of extremely rare configurations, it has been
possible to  show that the phase diagram of knots
formed by block copolymers is more complex than that of homolymers. New peaks appear in the heat
capacity corresponding to different transition processes.
Particularly interesting is the situation in which most of the monomers are subjected to repulsive interactions apart form a small number
of monomers that are  able to form contacts with each other. At the lowest temperatures,  such knots are found in compact conformations which get soon destroyed upon heating leading to a fast expansion of the knot. With growing temperatures, this expansion continues at a lower pace in Setup I. Finally, at high temperatures these knots behave as their homopolymer counterparts in a good solvent.
In some longer knots,
metastable compact states have been observed at low temperatures, signalising that knots can be subjected to relevant rearrangements of their structure
when heated. 
%arrived here on 20.07.2022. Add a more extended description of the obtained results.

The material presented in this paper is organised as
follows.
In Section~\ref{method} the used methodology is briefly explained.
The obtained results are discussed in Section~\ref{thermalprop}.
The thermal properties of knots in Setup I and Setup II are presented separately in Subsections \ref{thermalsetupI} and \ref{thermalsetupII} respectively.
Finally, the conclusions and open problems are the subject of Section~\ref{conclusions}.

\section{Methodology}\label{method}
Polymer rings are modeled as self avoiding loops on a simple cubic lattice. Monomers are located on the lattice sites and each lattice
side can be occupied by at most one monomer. Two consecutive monomers on the loop are linked by one lattice bond,
so that the total length of the knot in lattice
units is equal to $N$.
The energy  of a given knot conformation $X$ is expressed in Setup I
and Setup II by the 
following Hamiltonians respectively:
\begin{equation}
  H_I(X)=\varepsilon(m_{AA}+m_{BB}- m_{AB})
  \,\,\,\,\,\,\,\,\,\,\,\,\,\,\,\mbox{Setup I} 
  \label{hamI}
\end{equation}
and
\begin{equation}
  H_{II}(X)=\varepsilon(+m_{AA}-m_{BB})
  \,\,\,\,\,\,\,\,\,\,\,\,\,\,\,\mbox{Setup II}\label{hamII}
\end{equation}
In Eqs.~(\ref{hamI}) and (\ref{hamII}) the quantities
$m_{MM'}$'s count the numbers of contacts between monomers of the kind
$M$ and $M'$, where 
$M,M'=A,B$.
Let
$\bos R_1,\ldots,\bos R_N$ denote the locations of the $N$ monomers.
Two monomers $i$ and $j$ are said to be in contact if
$i\ne j\pm1$ and $|\bos R_i-\bos R_j|=1$.
$\varepsilon>0$ is an energy scale measuring the cost of one contact,
which can be positive or negative depending on the setup and on the
monomer types. We note that the Hamiltonian $H_{II}(X)$ of setup II is a variation of the HP protein model \cite{dill} with the $B$ monomers being identified with the polar (P) aminoacid residues. The difference is that in setup II we have that the $A$ monomers are repelling themselves due to the short-range interaction $+\varepsilon m_{AA}$, while in the HP model they are only subjected to excluded volume interactions.
For convenience, we will introduce the rescaled temperature
$\bos T=\frac {k_BT}\varepsilon$.
%To be more precise, $\bos T$ can be
%considered a rescaled temperature after choosing
%thermodynamic units in which the Boltzmann constant
%is equal to one. In these units the temperature $\theta$ is related to
%the usual temperature $T$ by the relation: $k_BT=\theta$.
 To go back from $\bos T$ to the usual temperature $T$ measured in Kelvins some assumptions on $\varepsilon$
are needed.
For instance, we suppose that the
strength $\varepsilon$ of the interactions is a multiple of the energy
associated with thermal fluctuations at
room temperature $T_0$, i.~e. $\varepsilon=qk_BT_0$, where $T_0\sim
298K^{\circ}$ and $q$ is a positive real constant. At this point
it is easy to see that the temperature $T$ is expressed in terms of $\bos T$ as follows:
$q\bos T T_0=T$. For example, if $q\sim1.5$, the point $\bos T=1$
corresponds to the temperature $T=1.5T_0\sim 447K^{\circ}$.
After the passage $T\longrightarrow \bos T$, it is possible to eliminate the $\varepsilon$ factor in the Hamiltonians of Eqs.~(\ref{hamI}) and
(\ref{hamII}). The upshot is that we obtain the following rescaled Hamiltonians: $\bos 
H_{I,II}(X)=\frac{H_{I,II}(X)}\varepsilon$.
Here
$H_{I,II}$ can be either of the two Hamiltonians defined in
Eqs.~(\ref{hamI}) and (\ref{hamII}).

The simulations are performed using the Wang-Landau Monte Carlo
algorithm \cite{wl}.
The initial knot conformations are obtained by elongating
the existing conformations of minimal length knots \cite{rechnitzer,sharein} until the desired final length is attained.
Knots up to six crossing according to the Rolfsen table are studied, though there is no restriction against including more complicated knots.
The details on the sampling and the treatment of the
topological constraints can be found in
Refs.~\cite{yzff} and \cite{yzff2013}.
The random transformations that are necessary for sampling the different knot conformations are the pivot moves of Ref.~\cite{madrasetal}.
In order to preserve the topological state of the system,  the pivot algorithm and excluded area (PAEA) method of Ref.~\cite{yzff} is applied.

The partition function of the polymer knot is 
given by:
\begin{equation}
Z(\bos T)=\sum_{ E= E_{min}}^{ E_{max}}e^{- E/\bos
  T}g(E)
\end{equation}
where $g(E)$ denotes the density of
states:
\begin{equation}
g(E)=\sum_X\delta(\bos H_{I,II}(X)-{E})
\end{equation}
$g(E)$ is
the quantity to be evaluated via Monte Carlo methods. $ E_{min}$
and $ E_{max}$ represent respectively the minimum and maximum
values of the energy. The whole energy range ${\cal I}=[
    E_{min}, E_{max}]$ over which the   
sampling is 
performed depends on the used setup, the length of the knot, its topology and the selected monomer distribution.
To determine the values of $E_{min}$ and $E_{max}$, a preliminary run  without specifying any energy limit is performed. In doing that we exploit the fact that
the Wang-Landau algorithm is very efficient in  exploring the whole energy range of the system. The preliminary run is stopped when no new values of the energy are found. After that, the averages of the observables are computed
by a second run with the values of $E_{min}$ and $E_{max}$ calculated from the preliminary run. Also in this second run the energy range is kept open, but
for the convergence of the Wang-Landau algorithm only the energy values in the interval $[E_{min},E_{max}$ are considered.
  For the convergence of the Wang-Landau algorithm the sampling of an order of $10^{12}$ conformations is necessary. If new values of $E_{min}$ and $E_{max}$ appear during the sampling, the run is repeated with the new, extended energy range. In the case of long polymers with $N\ge 300$, cuts in the energy range are necessary in order to obtain the corvergence of the Wang-Landau algorithm in a reasonable time. In this case, several runs are repeated by slightly changing the energy range to check that the results are independent of the energy range despite these small variations. It turn out that the Wang-Landau algorithm is very robust in this sense. For instance, small variations of the energy range do not have relevant influences on the height and the position of the peaks of the specific heat capacity. The averages of the observables are particularly insensitive under changes of $E_{max}$, while variations of a few percent occur at very low temperatures by changing $E_{min}$ in the case of block copolymers in setup I, which is the most critical with respect to the computational time.

The
expectation values of any observable $\cal O$ may be computed
using the formula:
\begin{equation}
  \langle{\cal O}\rangle(\bos T)=\frac 1{Z(\bos T)}\sum_{ E=E_{min}}^{E_{max}}e^{-E/\bos   T}g(E){\cal O}_{ E}
\end{equation}
\begin{figure}
  \begin{center}
    \includegraphics[width=0.48\textwidth]{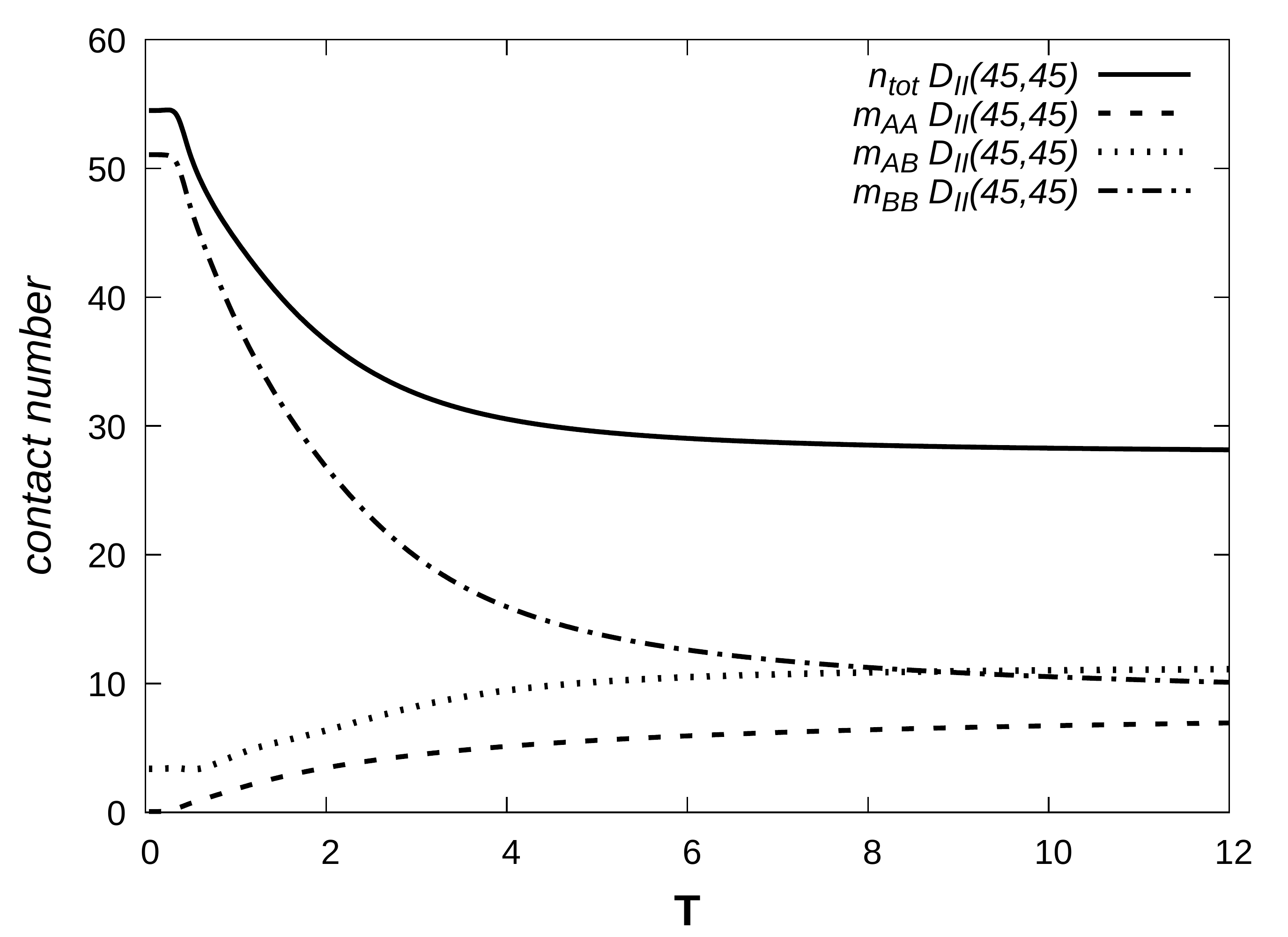}
    \includegraphics[width=0.48\textwidth]{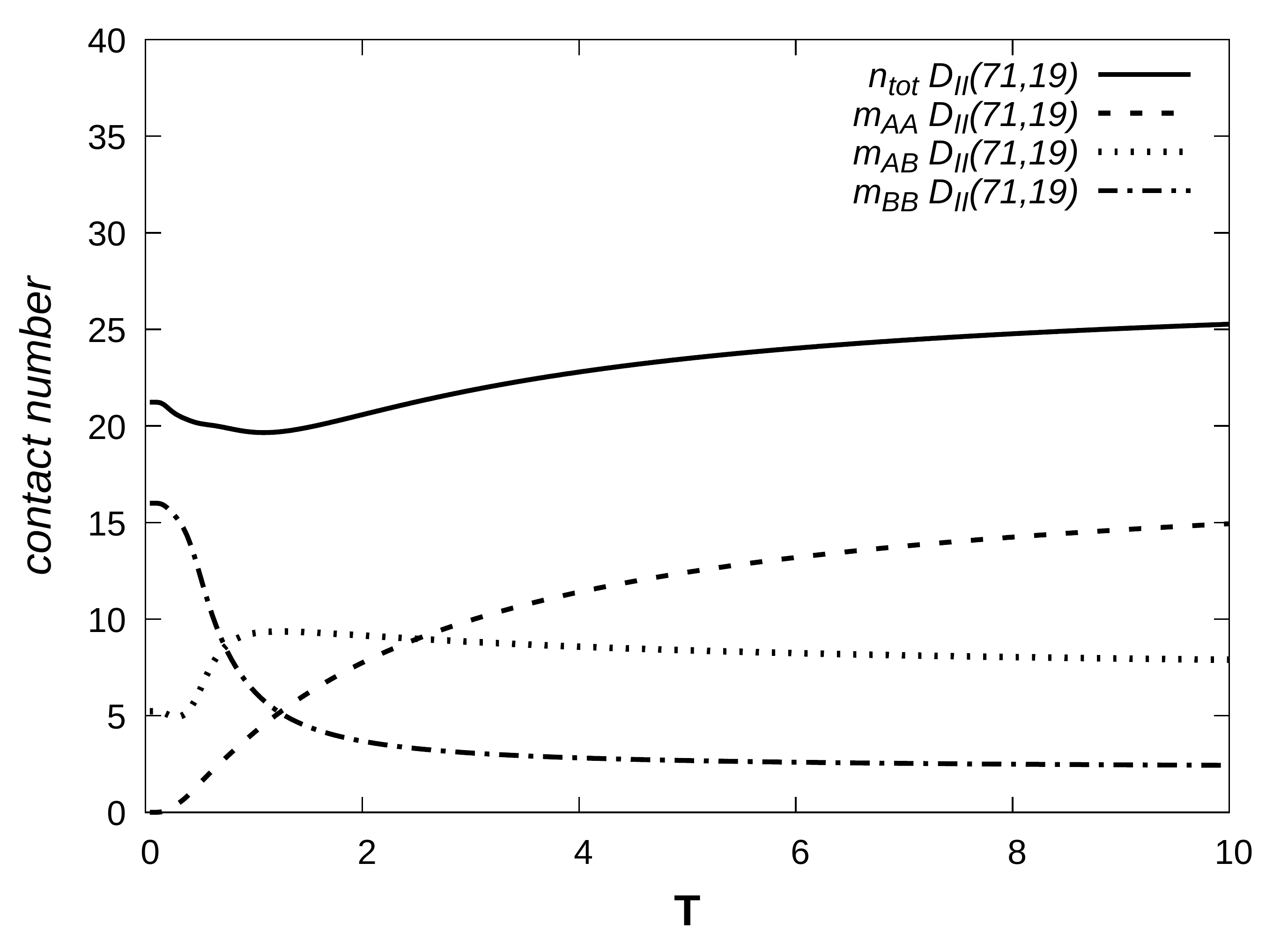}
    \caption{
While the specific energy is relevant for understanding the behavior of homopolymers, in the case of copolymers the numbers of contacts formed by the $A$ and $B$ monomers and the total number of contacts $n_{tot}=m_{AA}+m_{BB}+m_{AB}$ are very useful quantities too. The pictures show how the numbers $n_{tot},m_{AA},m_{AB}$ and $m_{BB}$ change with the temperature in the case of a knot $3_1$ with  monomer distributions $D_{II}(45,45)$ (left panel) and $D_{II}(71,19)$ (right panel). }\label{fig0a-ncont} 
  \end{center}
\end{figure}
%\begin{figure}
%  \begin{center}
%    \includegraphics[width=0.48\textwidth]{fig-typeII-E5.1-N90.pdf}
%  \caption{The specific energy of the knot $5_1$ with $N=90$ in two
%different monomer distributions $D_{II}(45,45)$ and $D_{II}(71,19)$ is plotted as a function of $\bos
%T$. }\label{fig0a} 
%  \end{center}
%\end{figure}
Here ${\cal O}_E$ denotes the average of $\cal O$ over all
sampled states with rescaled energy $E$. The observables that
will be considered in this work are the mean specific energy
\begin{equation}
\frac {\langle E(\bos
T)\rangle}N=\sum_{E= E_{min}}^{E_{max}}Ee^{-E/\bos
  T}g(E)
\end{equation}
the specific heat capacity $C/N=\frac
1N\frac{\partial \langle E(\bos T)\rangle}{\partial \bos T}$ and the
mean square 
average of the gyration radius $R_G^2$.
Additional information on the shape of the knot at different temperatures and energies has been gathered studying the number of contacts formed by the monomers and by closed inspection of the generated conformations.
\section{Thermal properties of knotted diblock copolymer
  rings}\label{thermalprop}
Both Setups I and II are characterised by the coexistence of the attractive and repulsive interactions. The monomers of type $A$ or $B$ may attract or repel themselves or the monomers of the other type. In Setup II only the excluded-volume forces are acting between the $A$  and $B$ monomers.
In this situation, depending on the temperature, a particular interaction can become more
relevant than the others in shaping the behavior of knotted block copolymers and lead to different regimes.
The averages of the numbers $m_{AA},m_{BB},m_{AB}$ of the contacts formed by the monomers of a specific type with themselves or with the other type together with the averaged total number of contacts $n_{tot}=m_{AA}+m_{BB}+m_{AB}$ provide some hint about the knot's conformations in these regimes.
An example of such plots is given in Fig.~\ref{fig0a-ncont}, in which a knot $3_1$ in Setup~II with different monomer distributions is considered.
Another quantity that is important is the specific energy. It turns out that, as it is expected, at any given temperature the specific energy of a copolymer knot
 never exceeds the specific energy of a homopolymer knot of equal length and
topology in a good solvent, where all monomers repel themselves.  On
the other side, it will always be bigger than that of a homopolymer in
bad solvents. 
%This feature can be seen in Fig.~\ref{fig0a}, where the data
%of the specific energy of a $3_1$ knot with $N=90$ segments are
%displayed. Indeed, in this figure the maximum and minimum values of
%the specific energy are 
% attained in the case of homopolymers in good and bad solvents
%respectively.
\begin{figure}
  \begin{center}
    \includegraphics[width=0.48\textwidth]{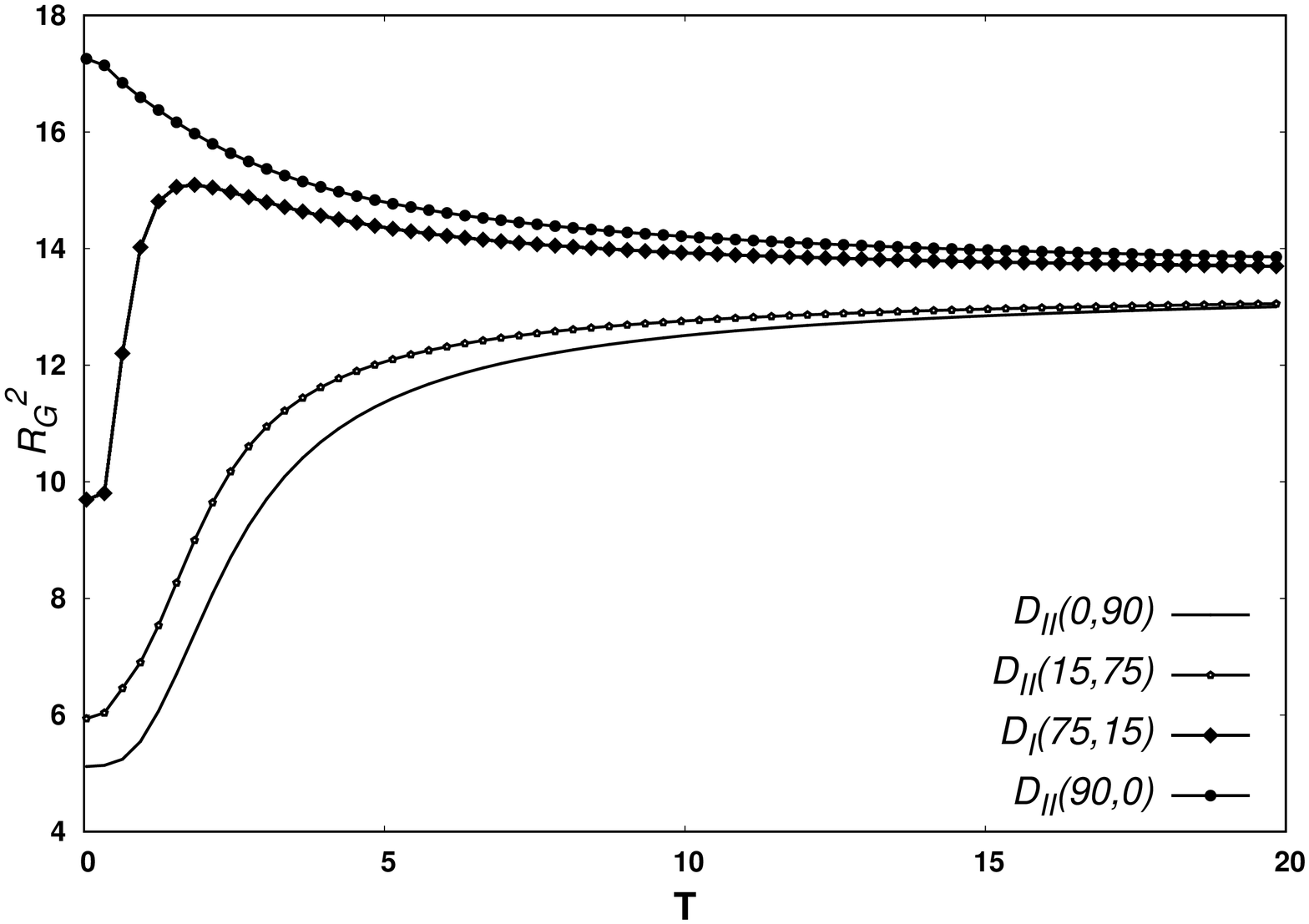}
  \caption{The gyration radius of the knot $3_1$ with $N=90$ in various
monomer distributions is plotted as a function of $\bos
T$. }\label{fig1a} 
  \end{center}
\end{figure}
A similar trend is observed in the case of the gyration radius as shown in  Fig.~\ref{fig1a}:
At any given temperature, the  gyration radii of  $3_1$ knots with
different monomer compositions
range between a minimum provided by the
the gyration radius of the $3_1$
homolymer knot $D_{II}(0,90)$  and an upper limit given by the $3_1$
homopolymer knot $D_{II}(90,0)$.
Another feature which is visible in Fig.~\ref{fig1a} is that
the swelling process under increasing temperatures may become much more rapid 
when the knot contains monomers of different kinds. For example, the
gyration radius of the knot $3_1$ with monomer distribution
$D_I(75,15)$ changes faster than that of the homopolymer
$D_{II}(0,90)$ and the diblock copolymer $D_{II}(15,75)$.
We also note that the diblock
copolymer $D_I(75,15)$ exhibits a
swelling phase at lower temperatures followed by a mild shrinking phase at higher
temperatures. This behavior becomes stronger  with growing topological
complexity.
\begin{figure}
  \begin{center}
    \includegraphics[width=0.48\textwidth]{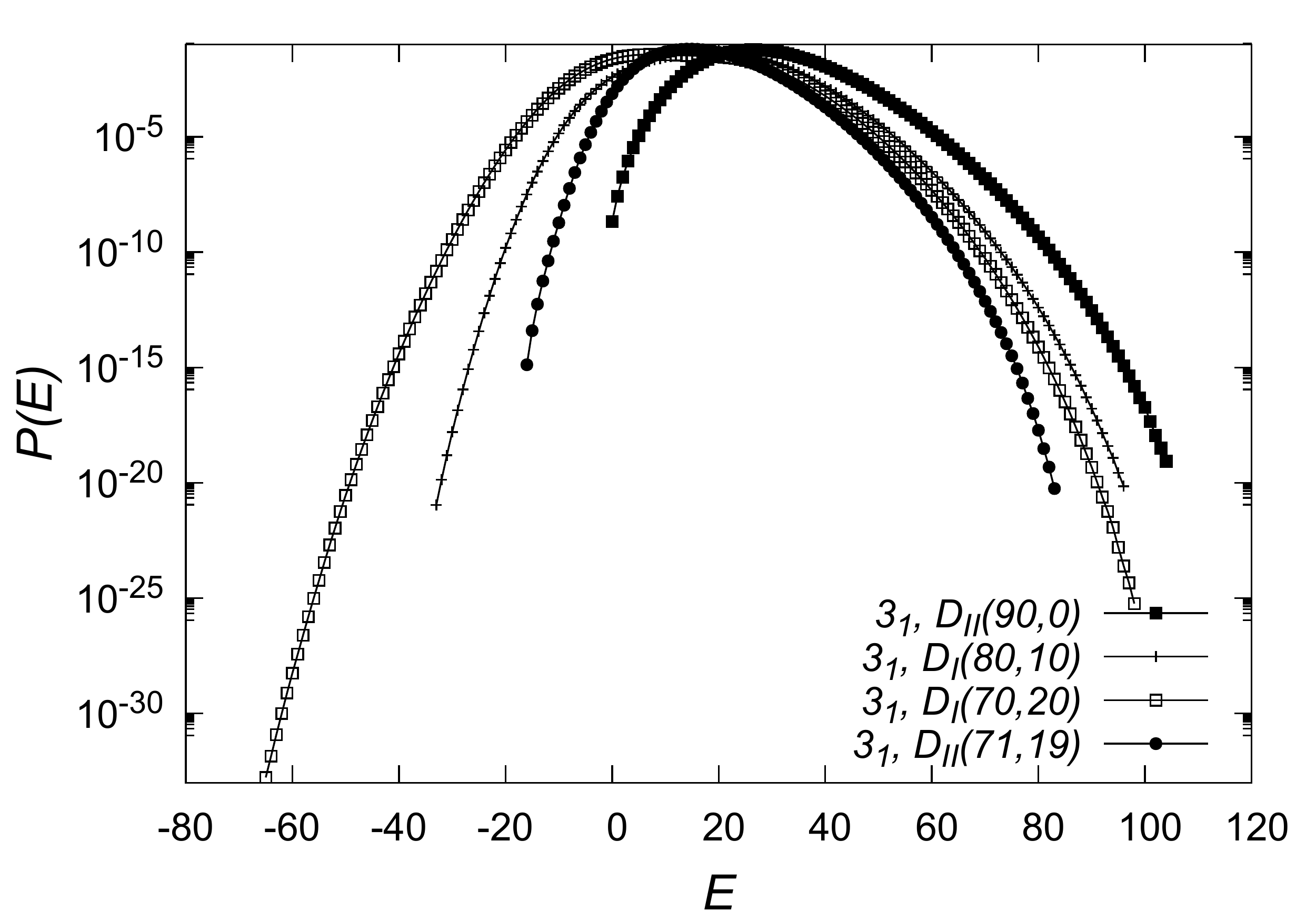}
    \caption{The probability $P(E)$ of obtaining a state of energy $E$
      for a knot $3_1$ with $N=90$ in various
monomer distributions is plotted as a function of the
energy $E$. }\label{fig-prob} 
  \end{center}
\end{figure}
In the following, we will discuss the results obtained for knots in
Setup I and Setup II separately.

To conclude the discussion on the general features of knots
formed by copolymers, we stress
that the Monte Carlo sampling is much more difficult for knots  
in Setup I than
in Setup II.
The reason can be understood by looking at
Fig.~\ref{fig-prob} where the probability $P(E)$ of states of energy
$E$  has been computed for a few knots in both setups.
That figure  shows that knots admit in Setup I 
conformations in the
lowest part of the energy spectrum that are much more rare than any knot conformation in Setup II.
This makes the sampling of the lowest energy states in Setup I very difficult, especially for long knots.
Of course, the energy spectrum, and thus also the existence of rare ultralow energy conformations, strongly depends on the monomer composition of a knot. This is visible by looking in Fig.~\ref{fig-prob}  at the differences between the same knot $3_1$ with $N=90$ in the monomer distributions $D_I(80,10)$ and $D_I(70,20)$.
However, even in the case in which the monomer distribution is fixed, the energy range grows considerably when passing from Setup I to Setup II.
 For example, the most compact state of
 a $3_1$ knot with $N=90$ and monomer distribution $D_I(70,20)$
has an energy $E<-60$ and
a
probability that is below $10^{-30}$. In the same knot $3_1$ with
$N=90$ and the similar monomer distribution $D_{II}(71,19)$, the probability
of the lowest energy state is higher of more than $15$ orders.
In a knot $4_1$ with $N=1000$ with monomer distribution  $D_I(800,200)$ the 
lowest energy state has a probability lower than $10^{-256}$.
\subsection{Results for Setup I}\label{thermalsetupI}
\begin{figure}
  \begin{center}
    \includegraphics[width=0.48\textwidth]{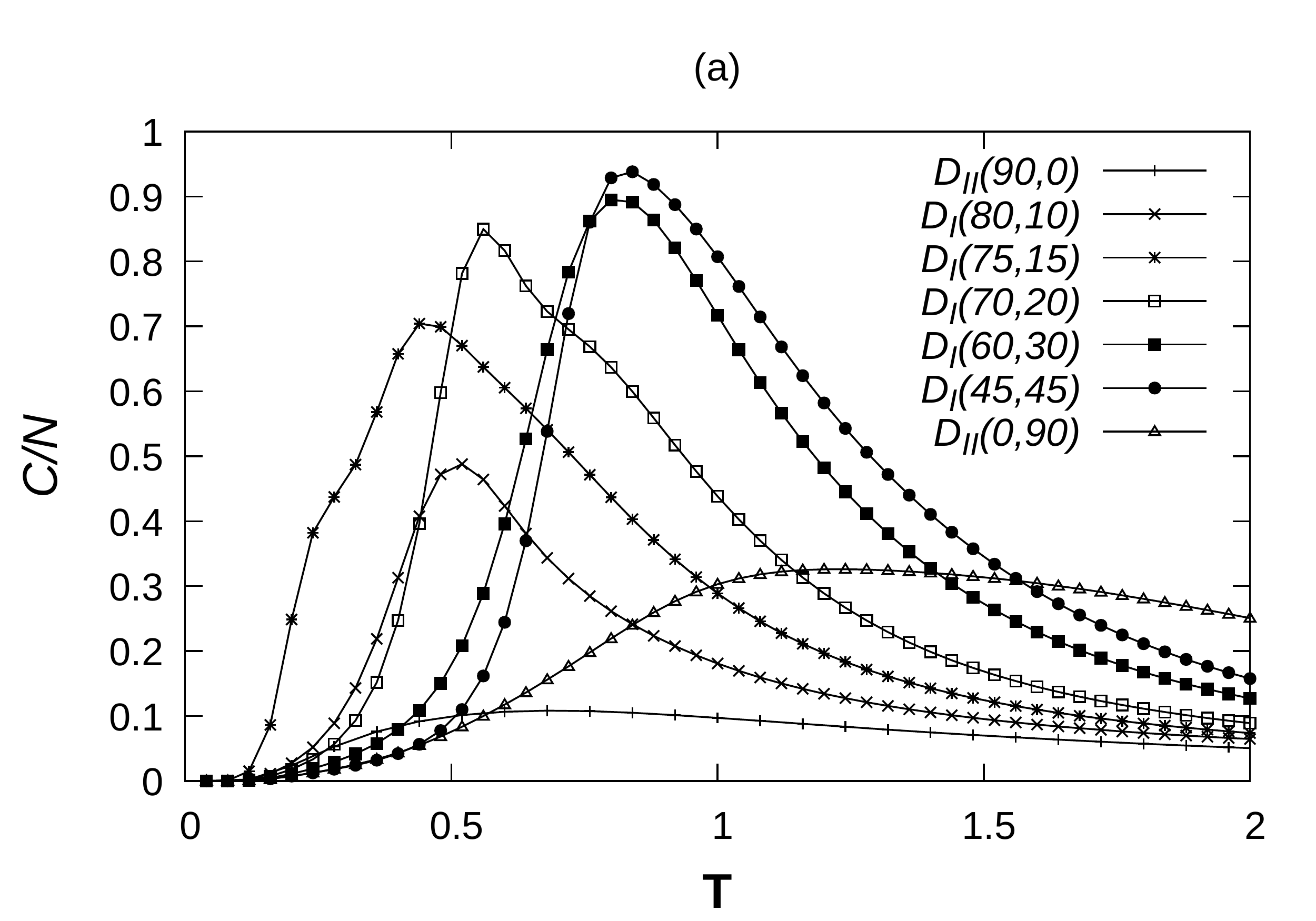}
    \includegraphics[width=0.48\textwidth]{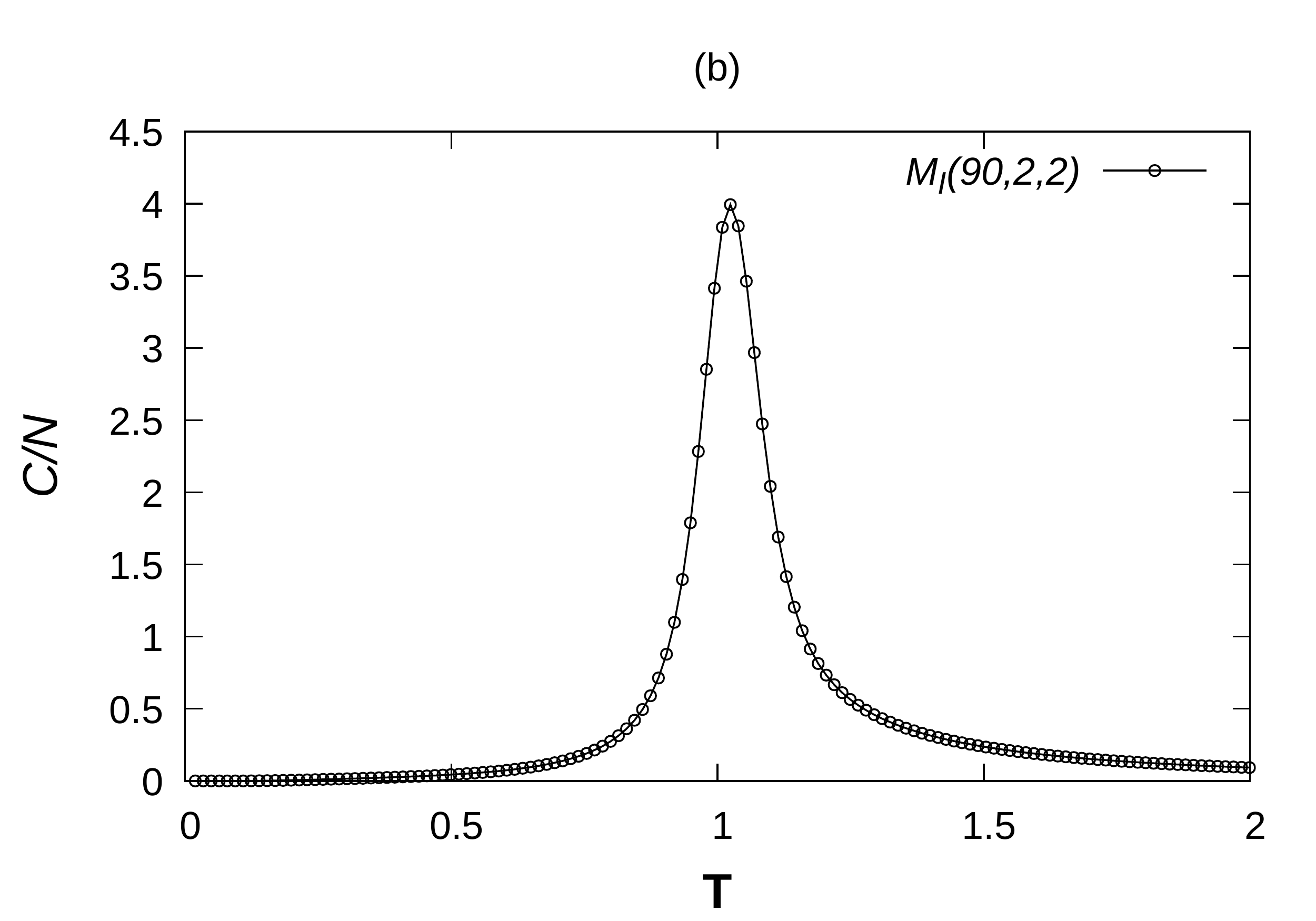}
    \caption{The specific heat capacity $C/N$ of a knot $3_1$ with $N=90$
   in various 
   monomer distributins is plotted as a function of $\bos T$.
The peak of $C/N$ is highest in the case of the monomer distribution
$M_I(90,2,2)$ (panel b)
and lowest in the case of the homopolymer $D_{II}(90,0)$
(plot with crosses at the bottom of the picture in panel a).
 }\label{fig-local-N90-b}
  \end{center}
\end{figure}

\begin{figure}
  \begin{center}
    \includegraphics[width=0.48\textwidth]{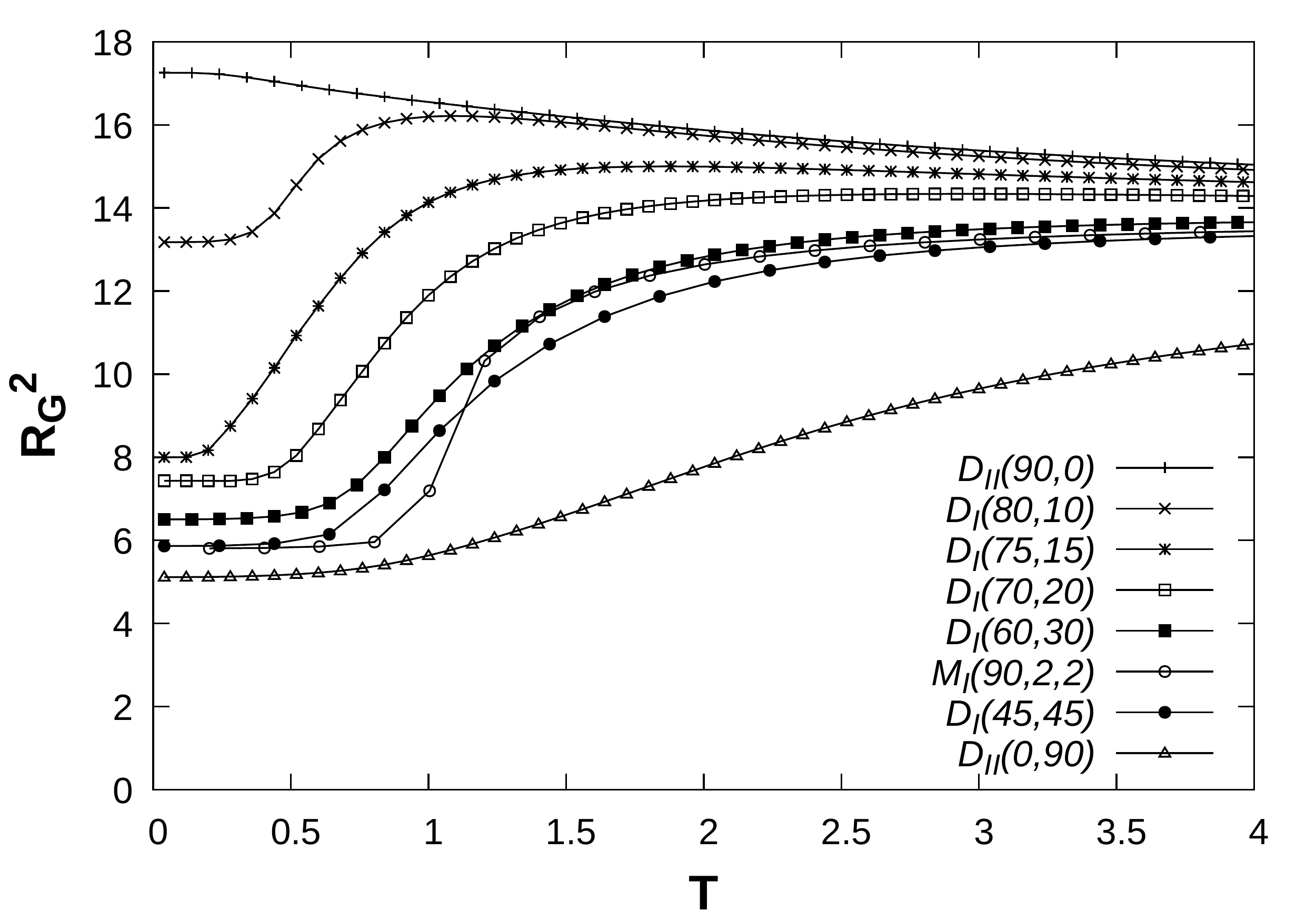}
    \caption{The mean square gyration radii $R_G^2$ of a knot $3_1$ with
      $N=90$ in various 
monomer configurations is plotted as a function of $\bos T$. Going
from the top to the bottom it is possible to distinguish the plots of
the gyration radii for the
following monomer distributions: $D_{II}(90,0)$, $D_I(80,10)$,
$D_I(75,15)$ and $D_I(70,20)$. Below, there is a partial overlapping
of the plots of the mean square
gyration radii for the distributions $D_I(60,30)$ (line with black squares),
$M_I(90,2,2)$ (line with white circles) and $D_I(45,45)$ (line with
black circles). The plot in the bottom part of the figure is that of the
homopolymer in a bad solvent 
$D_{II}(0,90)$ (white triangles). }\label{fig-local-N90-c}
\end{center}
\end{figure}

The variety of behaviors that it is possible to obtain in the case of
short copolymer
knots is shown 
in Figs.~\ref{fig-local-N90-b} and \ref{fig-local-N90-c}~\footnote{Let
  us note that in the present case the knot has length $N=90$. Due to
  the fact that $90$ is not divisible by four, the monomer sequence is
  obtained by joining together $\nu=22$ basic units of the kind $AABB$
  and putting at the end of the sequence  (monomers 89 and 90) two
  monomers of the $A$ type. As a consequence, the 
  sequence looks as follows: $AABB-AABB-\cdots-AABB-AA$.}.
Both
figures refer to the  same
trefoil knot $3_1$ of length 
$N=90$.
In Fig.~\ref{fig-local-N90-b} the range of temperatures has been
restricted to the interval $0.00\le\bos T\le 2.00$ in order to display
in details
the features of the peaks of the specific heat capacity $C/N$.
The different plots of the specific heat capacity and the mean
square gyration radius $R^2_G$ have been obtained only by varying the
distribution of the $A$ and  $B$ monomers.
It is possible to realize from Fig.~\ref{fig-local-N90-b} that the
 heights of the peaks of the specific heat capacity and the temperatures
 at which this peak occurs are very sensitive
to the momomer distribution.
The peaks' heights  range in fact from about $0.11$ for the
homopolymer $D_{II}(90,0)$ (see plot with crosses at the bottom of Fig.~\ref{fig-local-N90-b}, panel a, and
comments in the caption) up to
about $4.00$ in the case of the multiblock copolymer $M_I(90,2,2)$ in Fig.~\ref{fig-local-N90-b}, panel b.
In general, we observe that the swelling process when the temperature increases is much more abrupt in knots with monomer distributions $M_I(N,2,2)$
than in those with any other monomer distribution. Consequently, the specific heat capacities in knots with monomer distribution $M_I(N,2,2)$ are characterised by high and narrow peaks as it can be seen by comparing the plot of $C/N$ in Fig.~\ref{fig-local-N90-b}, panel b for the distribution
$M_I(90,2,2)$ with the plots of the other distributions
in panel a.
By gradually increasing the size of the basic unit in the multiblock copolymer, for instance choosing $M_I(90,4,4)$,
$M_I(90,8,8)$ etc., the peak of the heat capacity becomes gradually lower and 
 wider.
We also note that 
the temperature at which the peak of the specific heat appears can be
fine-tuned by choosing the monomer distribution. 
The temperatures and the height of the peaks related to different
distributions are displayed in Table~\ref{TableI}.
\begin{table}
\centering
\caption{In the second column of this table are reported the values
  of the height of the 
  peaks of the specific heat capacity for a knot $3_1$ with different
  monomer distributions. In the third column it is shown the
  temperature $\bos T_{MAX}$,
at which the heat capacity is at its maximum. The plots of the
specific heat capacity are displayed in Fig.~\ref{fig-local-N90-b}.}
\label{TableI}
\begin{tabular}{ccc}
\toprule\\
  Monomer\hspace*{1cm}& Peak\hspace*{1cm}& Temperature\\
  distribution\hspace*{1cm}& height  \hspace*{1cm}&  $\bos T_{MAX}$ \\\toprule
 $D_{II}(0,90)$\hspace*{1cm}& 0.33\hspace*{1cm} & 1.22  \\
 $M_I(2,2)$\hspace*{1cm}&3.99\hspace*{1cm}  & 1.03 \\
 $D_I(45,45)$\hspace*{1cm}& 0.87\hspace*{1cm} & 0.94   \\
 $D_I(60,30)$\hspace*{1cm}&0.72 \hspace*{1cm} &0.94   \\
$D_{II}(90,0)$\hspace*{1cm} &0.11\hspace*{1cm}  & 0.69  \\
$D_I(70,20)$\hspace*{1cm} &0.85\hspace*{1cm}  & 0.56\\
  $D_I(80,10)$\hspace*{1cm}&0.49\hspace*{1cm}&0.51\\
  $D_I(75,15)$\hspace*{1cm}&0.70 \hspace*{1cm} & 0.48
\end{tabular}
\end{table}
The data in the table  are ordered according to decreasing
temperatures $\bos T_{MAX}$, whose values range in the wide interval $0.48\le
\bos T_{max}\le 1.22$.

%Another remarkable
%phenomenon appears in the plots
%of the quantity $R_G^2$ of Fig.~\ref{fig-local-N90-c}. While homopolymers are  simple
%systems whose size
%steadily increases (in bad solvents) or decreases (in good solvents)
%with growing temperatures, 
%the $3_1$ copolymers $D_I(75,15)$ ($f\sim 0.83$) and in particular
%$D_I(80,10)$ ($f\sim 0.89$) exhibit a more complex behavior. Their
%mean square gyration radius is smallest at low temperatures and  
%increases up to its maximum value at intermediate temperatures. After
%that, it starts to decrease and finally stabilizes
%to some value between the maximum and the minimum at high
%temperatures.
%The presence of three different
%regimes, compact, ultra swollen and swollen is strongly dependent on
%the monomer composition. Within Setup I, diblock copolymers
%with medium to low values of $f$  are characterized by only two regimes. This is for instance the case of the 
% $3_1$ knots with monomer distributions $D_I(45,45)$ ($f=0.50$) and $D_I(60,30)$ ($f\sim0,67$).
%Arrived here on 16.06.2022
The distribution of the $A$ and $B$ monomers greatly influences also the allowed
range of possible sizes as shown in Fig.~\ref{fig-local-N90-c}. For example,
in
the $3_1$ copolymer knot in the setup
$D_I(80,10)$  the values of the mean square gyration radius are
restricted to the narrow interval $13\le R_G^2\le 16$. By passing to
the $D_I(75,15)$ distribution, a change that requires just the
substitution of  five monomers of type $A$  with monomers of type $B$,
the new range in which $R_G^2$ can take its values is $8.86\le
R_G^2\le14.89$. Thus even little variations in the monomer distribution are
able to introduce significant changes in the expectation values of the gyration radius. 
Let us note that all gyration radii 
in Fig.~\ref{fig-local-N90-c} converge to a common limit at very high
temperatures as it is expected, because at high temperatures the
interactions between the monomers are no longer relevant due to the
strong thermal fluctuations.  
From Fig.~\ref{fig-local-N90-c} it turns also out that, as expected, at any temperature
the self-attracting homopolymer $3_1$ knot 
$D_{II}(0,90)$ is always much smaller than all other trefoil
knots in which repulsive interactions are present.
 %\begin{figure}
%  \begin{center}
%    \includegraphics[width=0.48\textwidth]{fig-local-spec-Rg-N90-bis.pdf}
%    \caption{}\label{fig-local-N90-c-bis}
%\end{center}
%\end{figure}

 One goal of this work is to investigate
 how the topology of a knot influences its thermal  behavior.
In the case of homopolymers it is known that the topological effects
are particularly strong in short knots  and gradually fade out with
increasing length, see e.~g. \cite{yzff2013}.
This is true also in the case of copolymers.
\begin{figure}
  \begin{center}
    \includegraphics[width=0.48\textwidth]{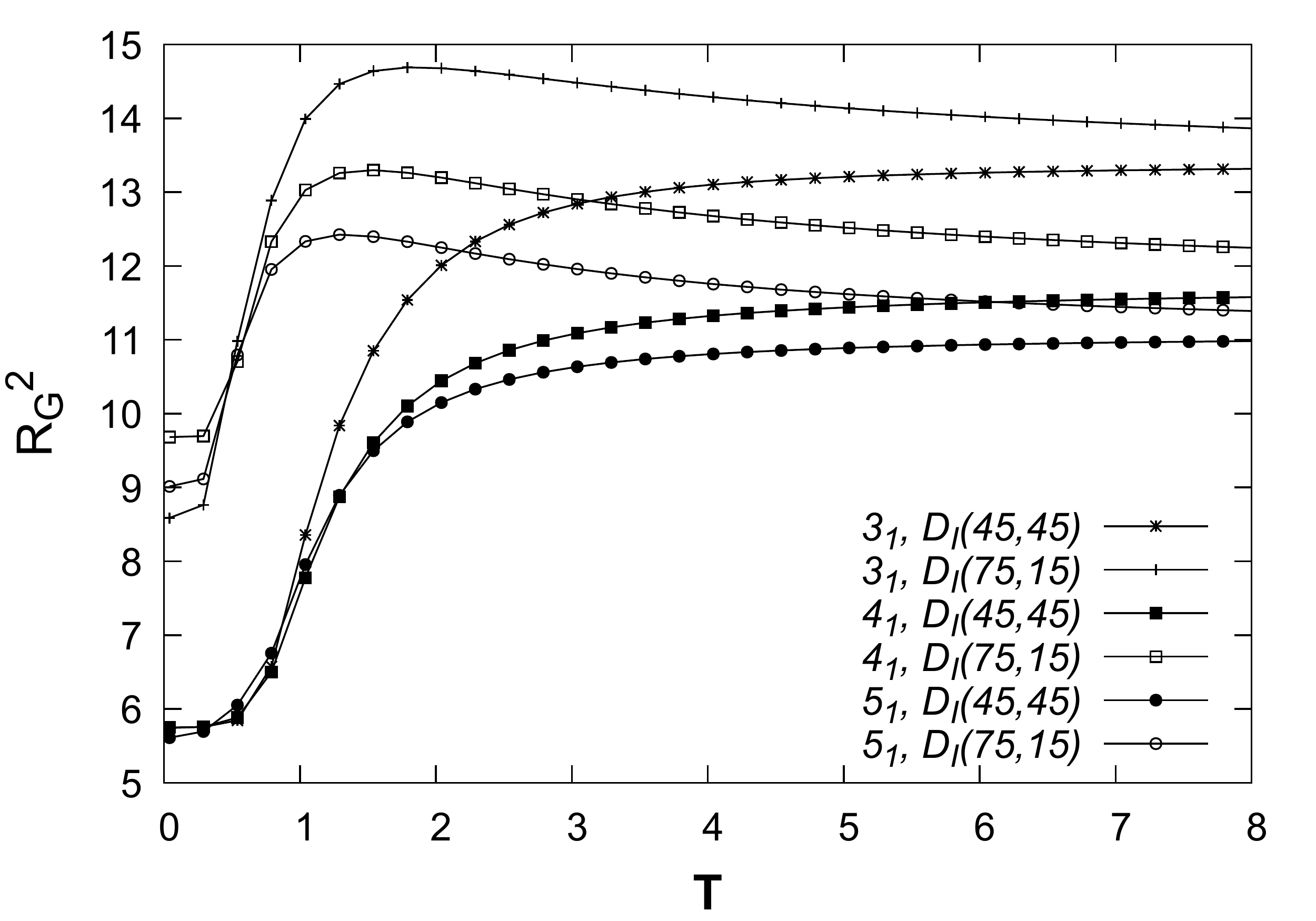}
    \caption{
Presented are the plots of $R_G^2$ in the case of the knots $3_1,4_1$
and $5_1$ with $N=90$. For each knot two values of $f$ are
considered: $f=0.50$, corresponding to the monomer
distribution $D_I(45,45)$ and $f\sim 0.83$, corresponding to the
monomer distribution $D_I(75,15)$.}\label{fig-top-N90}
\end{center}
\end{figure}
The plots in Fig.~\ref{fig-top-N90} show the gyration radii of
a few $AB-$diblock copolymers of knot types $3_1,4_1$ and $5_1$.
For each knot type two different monomer distributions have been
taken into account, namely $D_I(45,45)$ and $D_I(75,15)$.
%Arrived here on 08.05.2022
Fixing the monomer
distribution and the knot length, which is equal to $N=90$, we expect 
the differences in the plots to be due to pure topological effects. The
latter are quite evident in the figure. For instance, by looking at the
gyration radii of the knots with monomer distribution 
$D_I(45,45)$, it is clear that the sizes of these knots are
changing with the topology. The same conclusion is valid if we look at
the plots with  monomer
distribution $D_I(75,15)$.
It should be noticed that topological effects are less marked than
those connected with the modification of the monomer distribution.
This fact is well illustrated also by the plots of the specific heat
capacity. Fig.~\ref{fig-top-shc-N90} shows that by changing
the topology of a knot while keeping its monomer distribution fixed,
it is possible to shift  significantly the
peaks of the specific heat capacity but to a smaller extent than in
the case in which the topology is fixed
and the monomer distribution is modified.
\begin{figure}
  \begin{center}
    \includegraphics[width=0.48\textwidth]{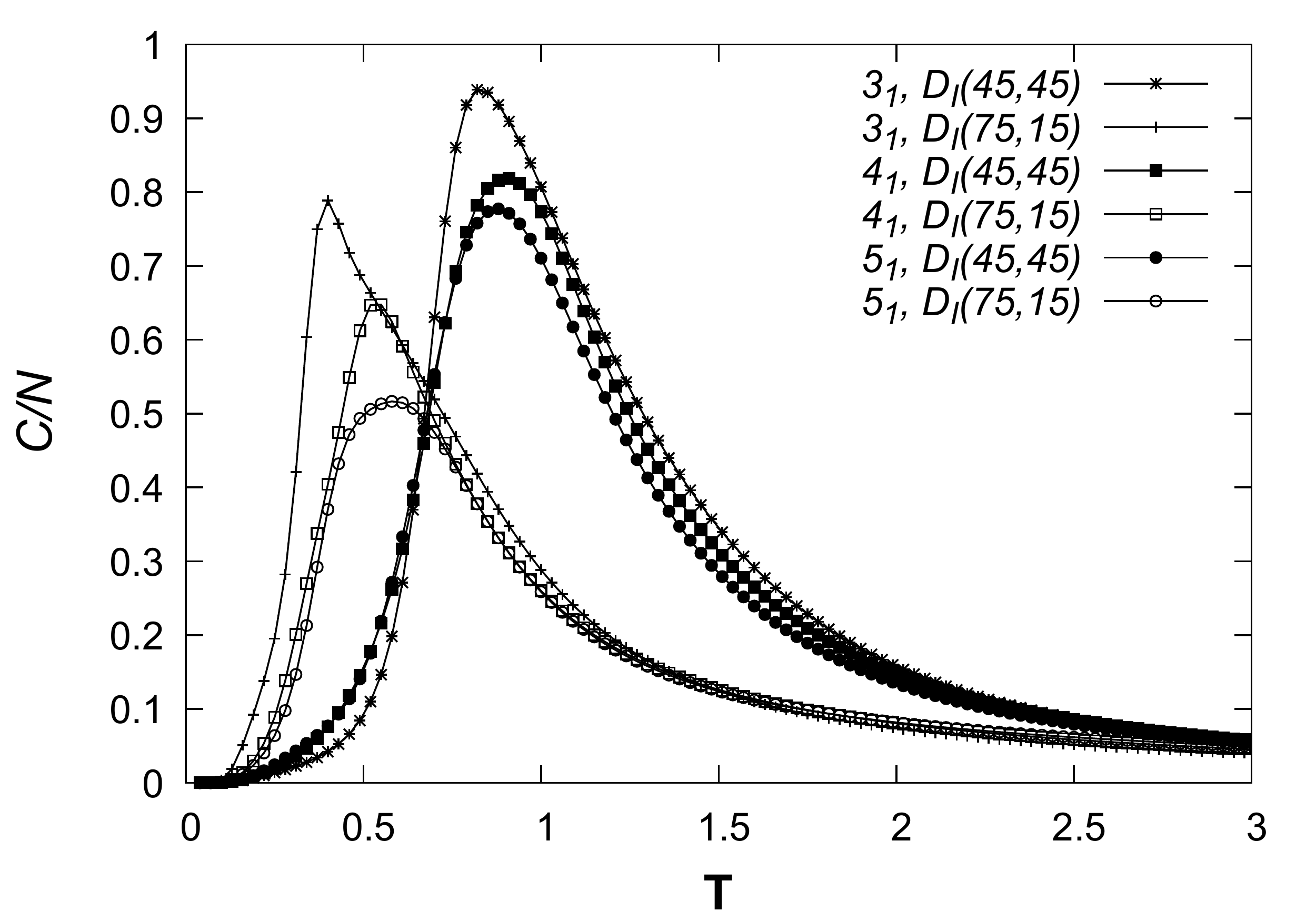}
    \caption{Dependence of the specific heat capacity $C/N$ on the
      topology and on the monomer composition $f$ for $AB-$diblock
      copolymers. 
Presented are the plots of $C/N$ in the case of the knots $3_1,4_1$
and $5_1$ with $N=90$. For each knots two values of $f$ are
considered: $f=0.50$ (case $D_I(45,45)$) and $f\sim 0.83$ (case
$D_I(75,15)$).}\label{fig-top-shc-N90} 
\end{center}
\end{figure}

\begin{figure}
  \begin{center}
    \includegraphics[width=0.48\textwidth]{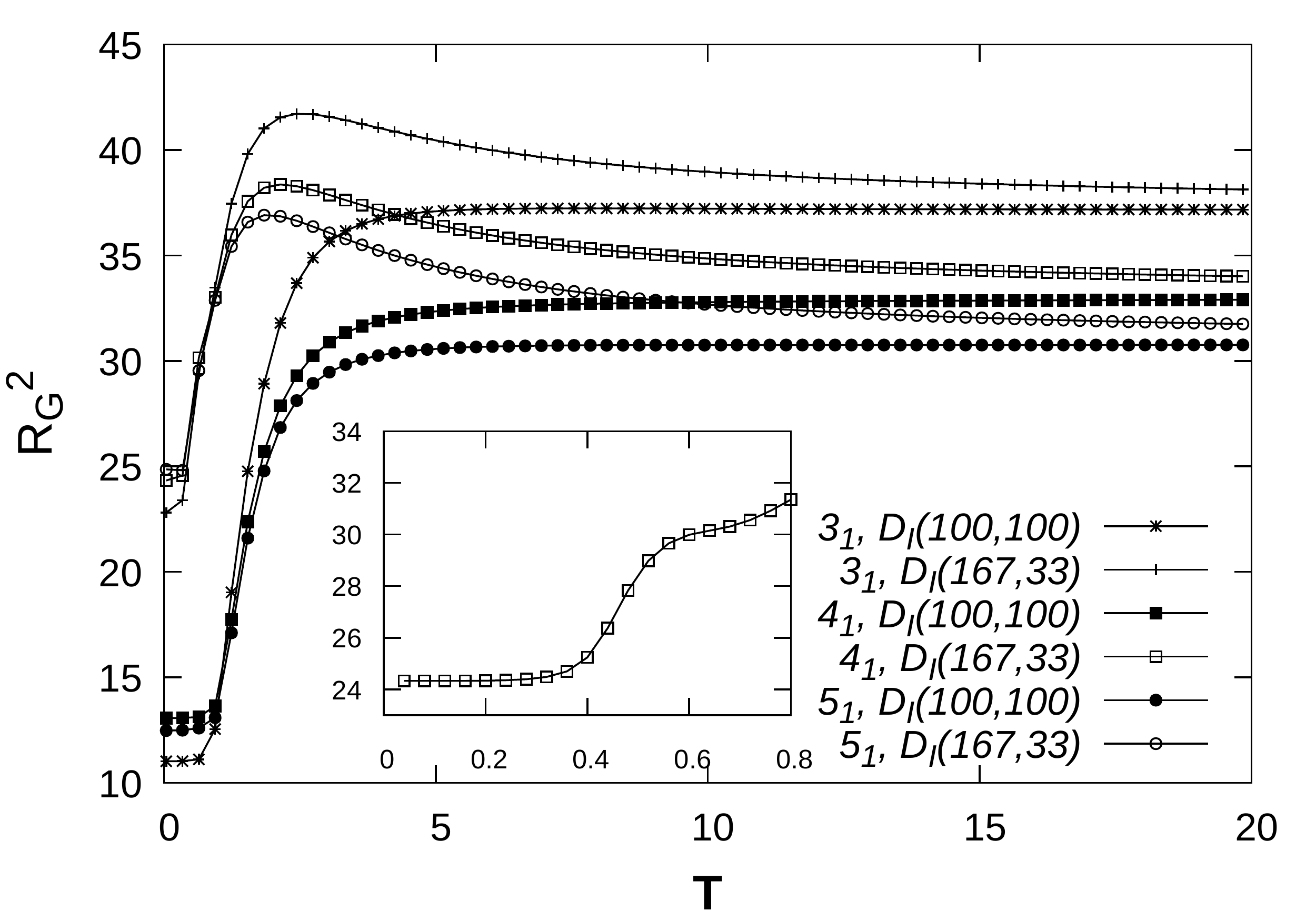}
    \caption{Dependence of the mean square gyration radius $R_G^2$ on the
      topology and on the monomer composition $f$ in the case of
      $AB-$diblock copolymers.
Presented are the plots of $R_G^2$ in the case of the knots $3_1,4_1$
and $5_1$ with $N=200$. For each knots two values of $f$ are
considered: $f=0.50$, corresponding to the monomer
distribution $D_I(100,100)$ and $f\sim 0.83$, corresponding to the
monomer distribution $D_I(167,33)$. In the inset it is shown the detail
of the behavior of $R_G^2$ for the knot $4_1$ with monomer
distribution $D_I(167,33)$.}\label{fig-top-rg-N200}
\end{center}
\end{figure}

\begin{figure}
  \begin{center}
    \includegraphics[width=0.48\textwidth]{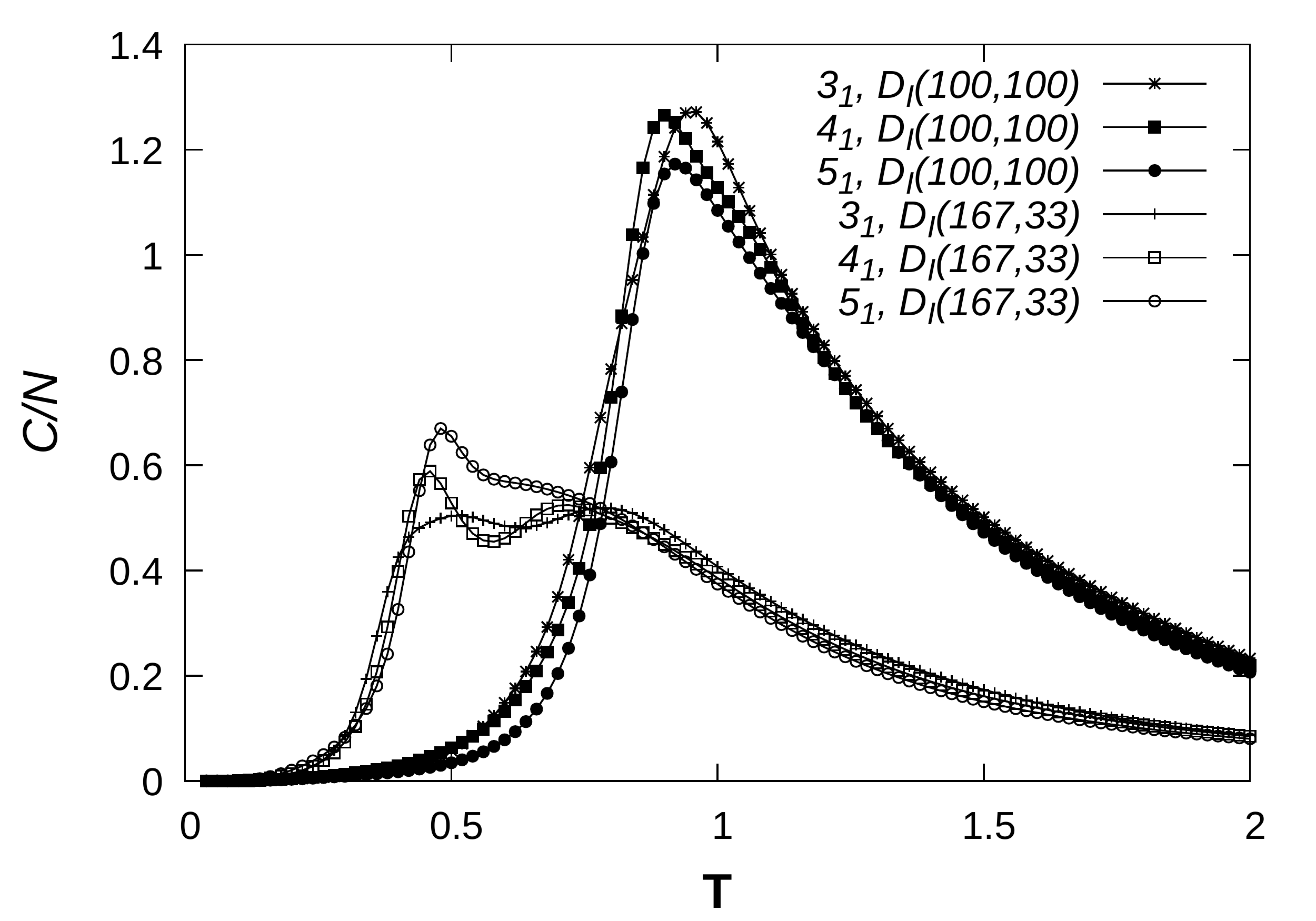}
\includegraphics[width=0.48\textwidth]{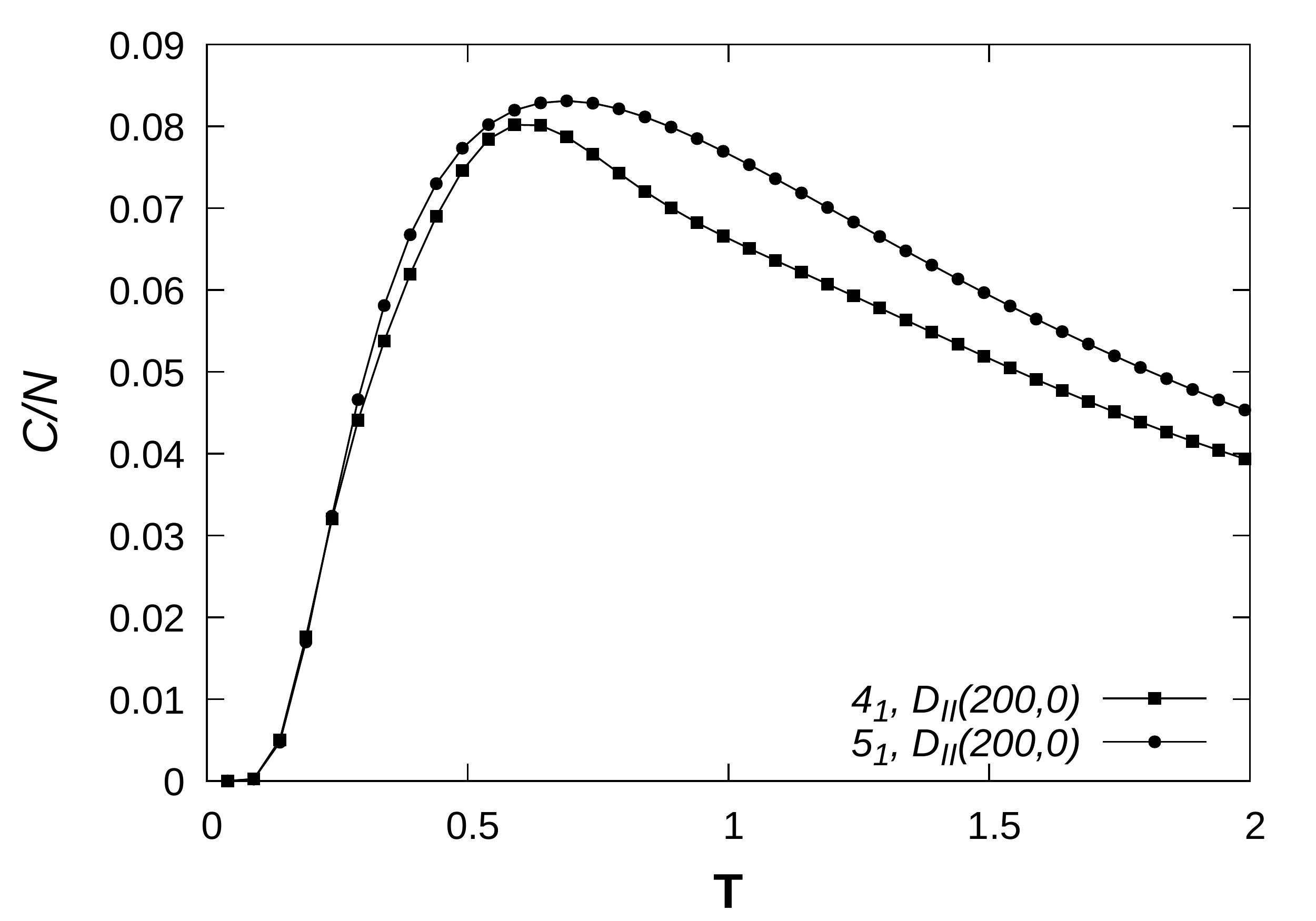} 
    \caption{The left panel shows the dependence of the specific heat
      capacity $C/N$ on the 
      topology and on the monomer composition $f$ for $AB-$diblock
      copolymers.  The plots of $C/N$ are presented in the case of the knots $3_1,4_1$
and $5_1$ with $N=200$. For each knot two values of $f$ are
considered: $f=0.50$ (case $D_I(100,100)$) and $f\sim 0.83$ (case
$D_I(167,33)$). The right panel shows the  specific heat capacity of
the homopolymer version of the knots $4_1$ and $5_1$ with $N=200$
fluctuating in a good solvent. 
    }\label{fig-top-shc-N200} 
\end{center}
\end{figure}

The above comments concerning the topological effects are valid also
in the case of longer polymers. 
The plots in
Figs.~\ref{fig-top-rg-N200} and~\ref{fig-top-shc-N200} of the
gyration radius and heat capacity respectively
illustrate the influence of both topology
and  monomer distribution on the thermal behaviour
of various knots of length $N=200$.
Fig.~\ref{fig-top-rg-N200} shows that topology is affecting
the knot size. Moreover, the shape of 
the peaks of the specific heat capacity changes with
the topology of the knot, as it is possible to see in Fig.~\ref{fig-top-shc-N200}. These changes cannot be attributed to statistical errors, as it has been verified by repeating the simulations starting from different seeds.
Of course, with increasing polymer lengths the topological effects
start to fade out.
In particular, the range of temperature in which the peaks of the specific heat capacity
are appearing  is approximately the same independently of topology for all
 knots considered
 in Fig.~\ref{fig-top-shc-N200}.
 This is not the case of the shorter knots with $N=90$, see
Fig.~\ref{fig-top-shc-N90}.

The novelty when $N=200$ with respect to shorter polymers
is the appearance of double peaks or of a peak with a shoulder in the specific
heat capacities of the knots with monomer 
distribution $D_I(167,33)$, see Fig.~\ref{fig-top-shc-N200}. A double peak
is characterising also the specific heat capacity of the knots
$0_1$ and $6_1$ with $N=200$ and monomer distribution $D_I(167,33)$ (not
shown in the figure).
This phenomenon
is accompanied by a  pattern that is visible in the
behaviour of the mean square gyration 
radius $R_G^2$ at low temperatures.
An example of this behaviour
is shown in the inset of
Fig.~\ref{fig-top-rg-N200}, in 
which the plot of $R_G^2$ of
a knot $4_1$ with monomer distribution $D_I(167,33)$ is displayed
in greater detail. As it is possible to see, there is a rapid
increase of $R_G^2$
in  the range of temperatures $0.4\le \bos
T\le 0.6$. This range coincides approximately with that in which the
peak of the specific heat
capacity of the knot $4_1$ centered at about 
$\bos T=0.50$ is appearing, see the line with empty squares in
Fig.~\ref{fig-top-shc-N200}.

%Arrived here on 11.05.2022
%The possibility that the double peaks or the shoulder in
%Fig.~\ref{fig-top-shc-N200}, left panel, could be related to the fact that
%monomers of type $A$ and $B$  could behave differently with growing %temperatures is to be   
%ruled out.
%If that would be true, we could expect
%in the
%specific heat capacity of the knots with $N=200$ and 
%monomer distribution  $D_I(100,100)$
%two similar peaks, one for the $A$ monomers and one for the $B$
%monomers.
%Instead,  the heat capacities of these knots are characterised by
%only one peak.

The appearance of double peaks and peaks with
shoulders  in the plots of the heat capacity of chains has been shown to be
related to the occurrence of
two different phase transitions in a small interval of
temperatures, see
\cite{rampf,janke1,janke2,proceedings} and \cite{velyaminov} in the case of polyelectrolytes. 
To understand what happens in the present context,
we recall the behaviour of homopolymers upon heating.
In homopolymers in a good solvent all monomers are subjected to purely repulsive interactions. At the lowest temperatures, the
number of  contacts between the monomers is minimal and the knot attains its maximally swollen size.
When the temperature is rising, the increasingly strong thermal fluctuations become energetic enough to break the potential barrier that prevents the  formation of contacts between the monomers. As a consequence, the size of the knot moderately shrinks with growing temperatures. This shrinking is moderate and the related process causes just a small peak in the heat capacity, see the examples of
Fig.~\ref{fig-top-shc-N200}, right panel.
In homopolymers in a bad solvent, on the contrary, the minimal size conformations are found when the temperature is very low.
The compact states that arise in this case consist of a large number of contacts. With growing temperatures, the number of contacts decreases and the knot continues to swell up to high temperatures until the thermal fluctuations dominate over the interactions. At this point, the gyration radius and the number of contacts of the ring are completely determined by entropy and the knot's topology. This swelling process in a bad solvent generates in the plots of the specific heat capacity a broad peak
extending over a large interval of temperatures.

In block copolymers, the situation is different because there are simultaneously both attractive and repulsive interactions.
Knotted diblock copolymers with $N_A\sim N_B$ have a behaviour similar to that of knotted homopolymers in  a bad solvent.
The reason is that, when $N_A\sim N_B$, the attractive forces are very strong because there is a large number of $A$ and $B$ monomers that may build contacts with each other. As a consequence, for energy reasons, at very low temperatures the number of contacts between the $A$ and $B$ monomers is at its maximum and the volume occupied in space by the knot is very small, though not so small as in the homopolymer case. With growing temperatures, similarly to what happens for knotted homopolymers in a bad solvent, the number of contacts decreases and the swelling process continues also when the temperature is high. In knots of length $N=90$ and $N=200$ the expansion  is completed at a temperature of $\bos T\sim 3$ when $N_A=N_B$.
The result is a broad peak in the specific heat capacity with the maximum of the peak at a temperature of $\bos T\sim 1$ (see Figs.~\ref{fig-top-shc-N90} and \ref{fig-top-shc-N200}) or higher like in the case of the knot $5_1$ with length $N=500$ and monomer distribution $D_I(250,250)$ of Fig.~\ref{fig-N500}.
%Arrived here on 30.07.2022
 A thorough investigation with a knot $4_1$ of length $N=200$ has shown that the behaviour of knots with $N_A\sim N_B$ described before still persists at least until $N_A=150$ and $N_B=50$. This means that the heat capacity is characterised only by a single peak. Of course, the heights of this peak will decrease with decreasing values of $N_B$ because less and less contacts between the $A$ and $B$ monomers are possible.

\begin{figure}
  \begin{center}
    \includegraphics[width=0.48\textwidth]{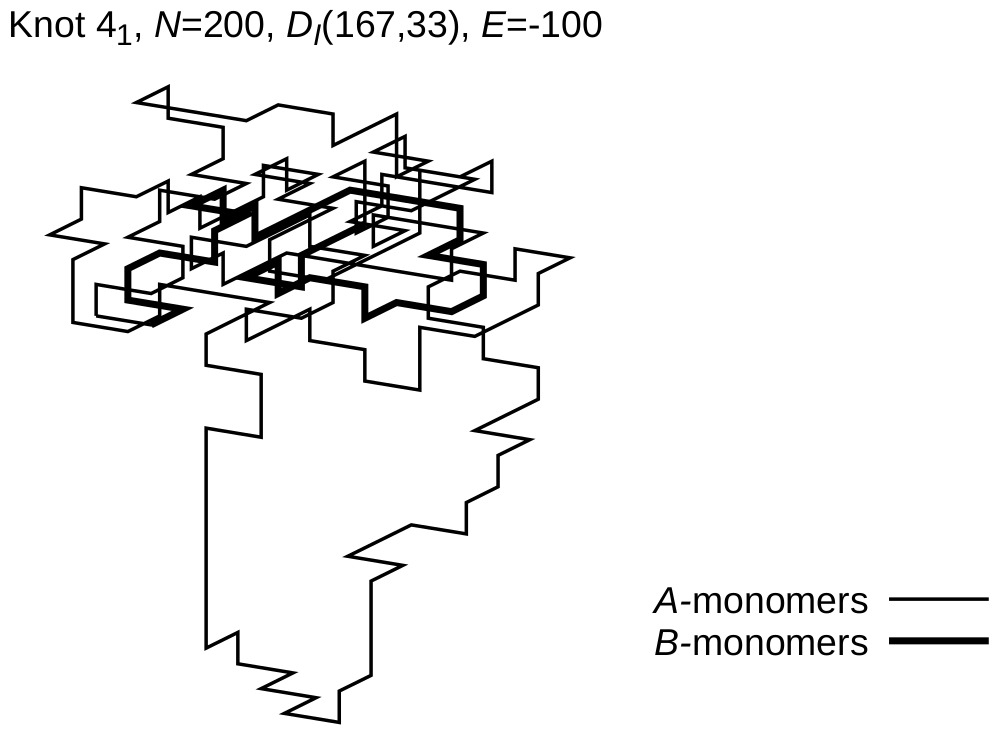}
    \includegraphics[width=0.48\textwidth]{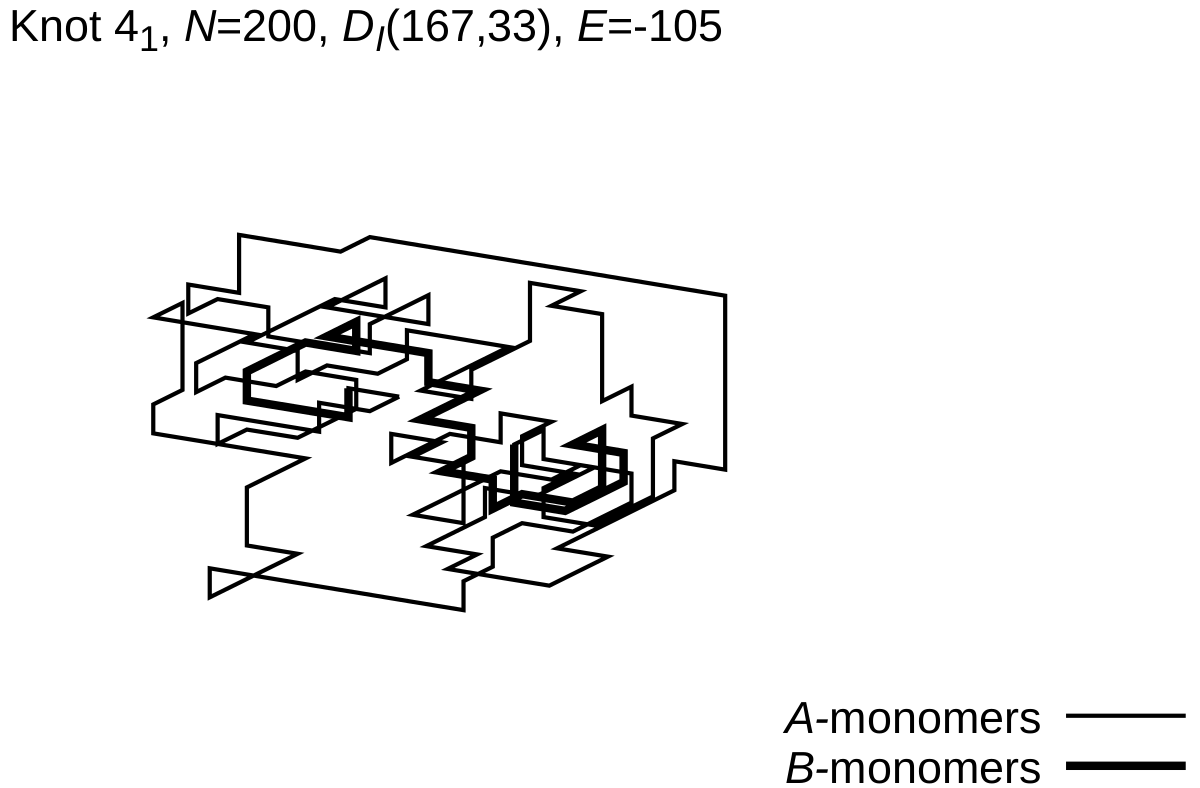}
      \end{center}
  \caption{  This figure shows two sample conformations of a knot $4_1$ with length $N=200$ and monomer distribution $D_I(167,33)$. The left conformation has energy $E=-100$. It is characterised by a partially ordered portion located in the upper part of the knot in which some of the $A-$monomers are in contact with the $B-$monomers. Other $A-$monomers form a "tail" on the bottom of the knot.
    At the just slightly higher energy of $E=-105$, the sample conformation of the knot appears  very different. The chain containing the $B-$monomers is more stretched than in the case $E=-100$. This allows the formation of two partially ordered portions of the knot concentrated at both ends of the chain with the $B-$monomers. In both pictures the knots have been rescaled in the same way to fit into the page and no different rescaling has been applied to different axes.}\label{comparison-115-103} 
\end{figure}

 Below a certain threshold of the number  $N_B$ of available $B$ monomers, knots depart significantly from the behaviour described above.
 To fix the ideas, we will say that such knots have 
monomer configurations of the type  $N_A>>N_B$.
In this case
the number of $B$ monomers that are able to form contacts with the $A$ monomers is extremely limited. This explains why the heights of the peaks of the specific heat capacity are decidedly lower than those of knots in which the monomer distribution is $N_A\sim N_B$, see Fig.~\ref{fig-top-shc-N200}.
Despite the fact that $N_B<<N_A$, at very low temperatures the number of contacts $m_{AB}$ between the  monomers of types $A$  and $B$ is quite high in order to minimise the energy according to the Hamiltonian~(\ref{hamI}).
Indeed, a knot $4_1$ with $N=200$ and monomer distribution $D_I(167,33)$ has been studied in details in order to understand the origin of the double peak when $N_A>>N_B$. It turns out that at the minimum energy value  $E_{min}=-111$, the average total number of contacts formed by this knot is $\langle n_{tot}\rangle=122$. The overwhelming majority of these contacts is due to the $A$ and $B$ monomers, because at $\bos T\sim 0$ we have that $\langle m_{AB}\sim 118$, $\langle m_{AA}\rangle\sim 4$ and $\langle m_{BB}\rangle =0$.
Of course, with rising temperatures $m_{AB}$ will decrease similarly as in the case $N_A\sim N_B$ illustrated before. The striking difference between the cases $N_A\sim N_B$ and $N_A>>N_B$ is however  that
knots with $N_A>>N_B$ undergo an extra rearrangement of their structures at very low temperatures.
More precisely, knots with $N_A>>N_B$ are approaching a very compact state when the temperature becomes very low like their counterparts with $N_A\sim N_B$.
In the case of the prototype knot $4_1$ mentioned above, this state contains  $\langle m_{AB}\rangle\sim 103$ contacts between the $A$ and $B$ monomers. The breaking of these contacts causes a swelling process that produces in the plot of the specific heat capacity a peak around the temperature $\bos T\sim 0.7$, see Fig.~\ref{fig-top-shc-N200}, left panel, plot with white circles.  This is the origin of the second peak in the heat capacities of the knots with monomer distribution $D_I(167,33)$ in Fig.~\ref{fig-top-shc-N200}.
%Arrived here on 31.07.2022
When the temperature further decreases, the quantity $R_G^2$ of the $4_1$ knot changes very rapidly from
$30$ to $24$, see the inset of Fig.~\ref{fig-top-rg-N200}.
This points to the fact that a rearrangement of the structure of the knot has taken place.
This rearrangment is connected with the formation of only a small number of new contacts.
Unfortunately, the close inspection of very rare conformations at the lowest energies
is very difficult, but it has been possible to capture conformations at the energies $E=-105$ and $E=-100$ of the knot $4_1$ with $N=200$ and monomer distribution $D_I(167,33)$, see Fig.~\ref{comparison-115-103}.
Basing on the available data, our interpretation of the phenomena associated with the presence of the two peaks in the specific heat capacity of knots in which $N_A>>N_B$ is as follows.
The second peak, appearing in all observed cases at temperatures $\bos T>0.5$, is due to the breaking of the contacts that are responsible for a bulk compact state that arises due to the attractive forces between the $A$ and $B$ monomers.
The first peak, observed at temperatures $\bos T< 0.5$, is due to a rearrangement of the knot, or at least a part of it,  involving the formation of just a few additional contacts. We suspect that the rearrangement is mostly concerning the segment with the $A$ monomers. Above the transition, there is a bulk compact state which is held together by the portion of the $A$ monomers that are in contact with the $B$ monomers. The other $A$ monomers cannot form contacts because of the few $B$ monomers available. As a consequence, these $A$ monomers form a long tail departing from the bulk compact state and fluctuating almost freely. Below thetransition, a rearrangement leading to the formation of additional contacts with more $A$ monomers becomes possible and the tail disappears or is reduced. 
%create a new picture called \label{comparison-115-103}
%E_{min}=-111 
%E_{min}'=-100
%Full
%cut-off E'_{min}
A close inspection of the conformations of the knot $4_1$ that have been captured confirms the above claim, see Fig.~\ref{comparison-115-103} and comments in the related captions.
%Arrived here on 03.08.2022

Other rearrangements could become possible in longer polymers.
They give rise to metastable states that may appear when the temperature is low, so that the formation
of contacts between the $A$ and $B$ monomers is energetically convenient.
Such metastable states and also the double peak are not observed in short knots,
like for instance those with $N=90$.
The main reason is that
in a short knot, the interactions between the monomers are more
frequent than in a longer one. As a consequence, the
rearrangements such that more monomers of the $A$ type will be in contact with monomers of the $B$ type will lead unavoidably also to contacts between the $A$ and $B$  monomers with themselves. This will increase the energy of the obtained conformation due to the repulsive interactions to which monomers of the same type are subjected.

One additional transition is related to a
process analogous to the shrinking taking place at high temperatures in the case of
homopolymers in a good solvent.
Indeed, since the $A$ monomers are numerous in the case $N_A>>N_B$ and they are
subjected to repulsive interactions,
we expect that, at high temperatures, the knot will slightly shrink as 
homopolymers do. This shrinking process causes in homopolymers just a small peak in the heat capacity, see
Fig.~\ref{fig-top-shc-N200}, right panel, so that in knotted diblock copolymers the effect on the plots of the heat capacity will be limited. However, the shrinking is visible in the plots of the gyration radius.

The general picture presented above fits very well with
the results obtained for knots of length $N=200$.
The plots of the specific heat capacity of
Fig.~\ref{fig-top-shc-N200}, left panel,
clearly show that the knots with monomer composition $f\sim 0.83$ ($N_A>>N_B)$ have
much lower peaks 
than those with $f=0.50$ ($N_A\sim N_B$). 
Moreover, the peaks of the specific heat capacity of the knots 
with $f\sim0.83$ occur at significantly lower  temperatures ($T\leq 0.50$ the first peak and $\bos
T\leq 0.70$ the second peak) than the
peak of the specific heat capacity of the knots with $f\sim 0.50$
($\bos T\sim 1.00$).
The sharp increase of the gyration radius following the 
%Arrived here on 18.06.2022-bis
Interestingly, out of all investigated knots $0_1,3_1,4_1,5_1$ and
$6_1$ with monomer composition $f\sim 0.83$ (the plots of $0_1$ and
$6_1$ are not reported), the second peak 
has been replaced by a
shoulder only in the heat capacity of the knot
$5_1$.

Let's now discuss the results of the knots with monomer distribution
$D_I(100,100)$. 
As previously discussed, in this case 
double peaks and shoulders are absent from
the plots of diblock copolymers when $N_A\sim N_B$. The plots with the
heat capacities of knots with $D_I(100,100)$ pointed out in
Fig.~\ref{fig-top-shc-N200} confirm this expectation.
Moreover, the peaks of the specific heat capacity are much higher than
those of knots with $D_I(167,33)$.

%arrived here on 18.06.2022-tris

\begin{figure}[h]
  \begin{center}
    \includegraphics[width=0.48\textwidth]{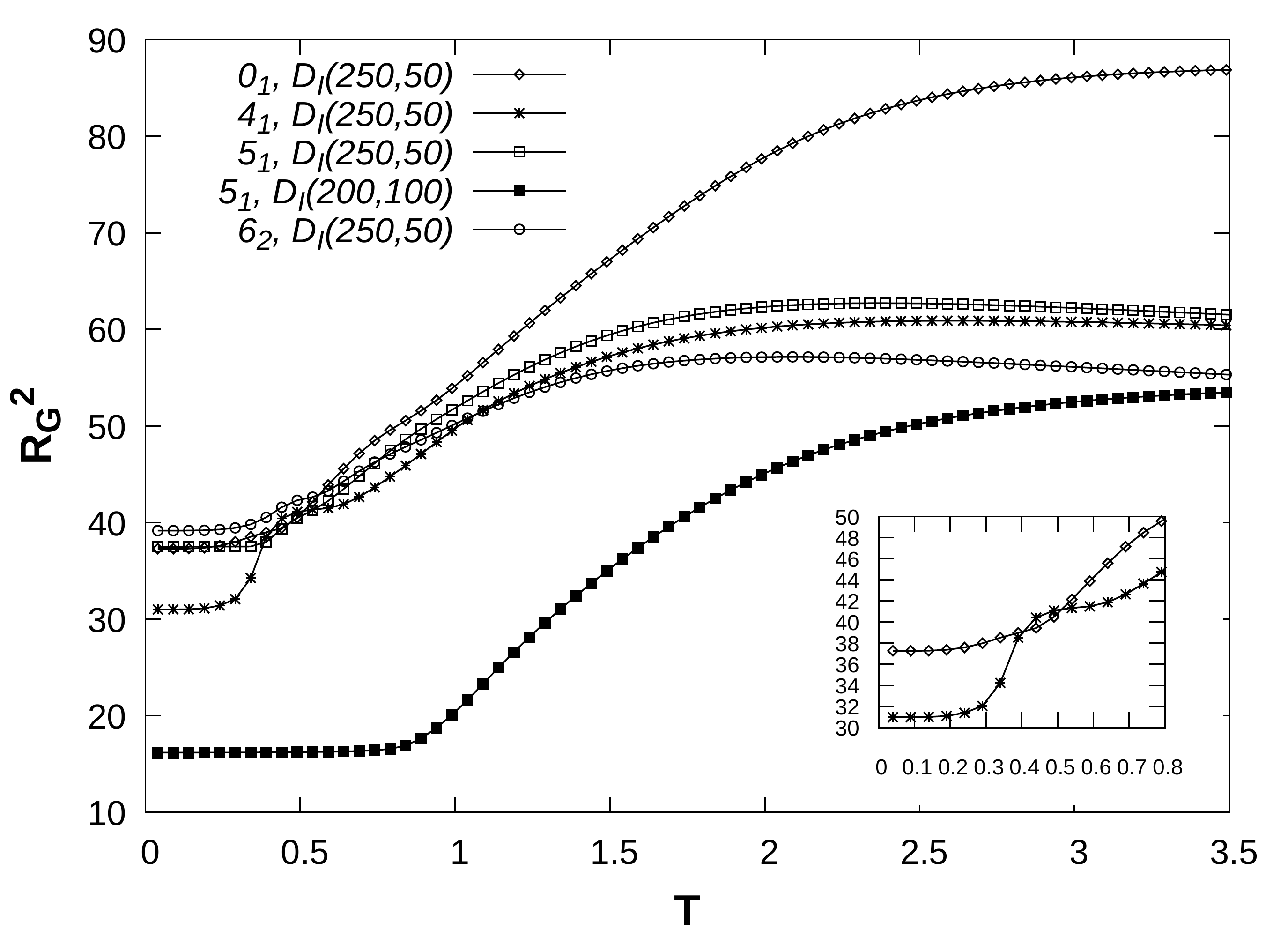}
    \caption{The data of the  gyration radius $R_G^2$ of knots $0_1$,  $4_1$, $5_1$ and $6_2$ of length 
      $N=300$ and monomer distribution
      $D_I(250,50)$ are presented. In the case of knot $5_1$ it is shown  the plot of $R_G^2$ also for the monomer distribution $D_I(200,100)$. In the inset the behaviour of the gyration radius of knots $0_1$ and $4_1$ is displayed in more details at low temperatures. Note the characteristic saddle point in the plot of the gyration radius of knot $4_1$ at  $\bos T\sim 0.45$.\label{fig-longer-shc}}\end{center} 
\end{figure}

\begin{figure}
  \begin{center}
    \includegraphics[width=0.48\textwidth]{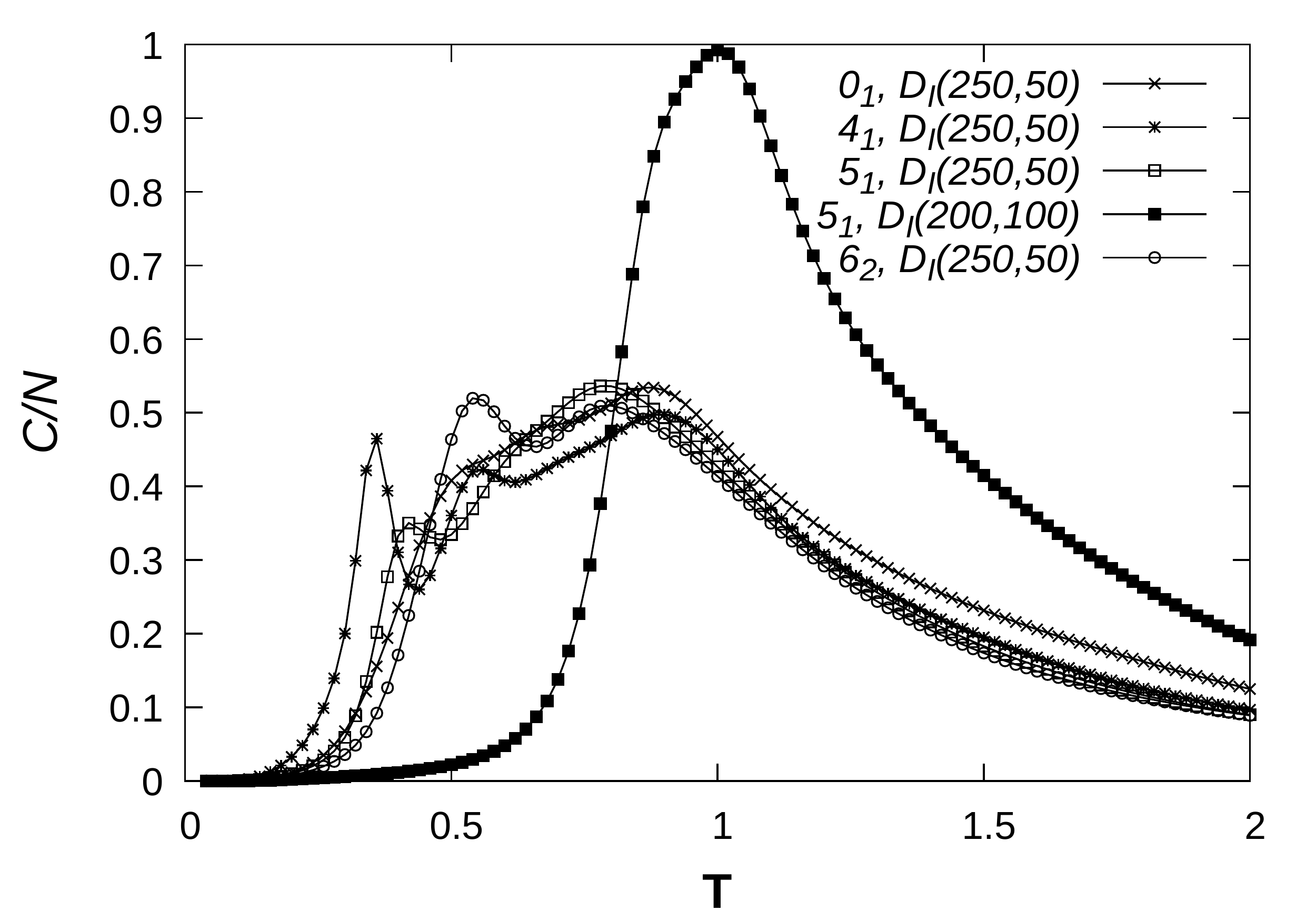}%{fig-new-events.pdf}
 \includegraphics[width=0.48\textwidth]{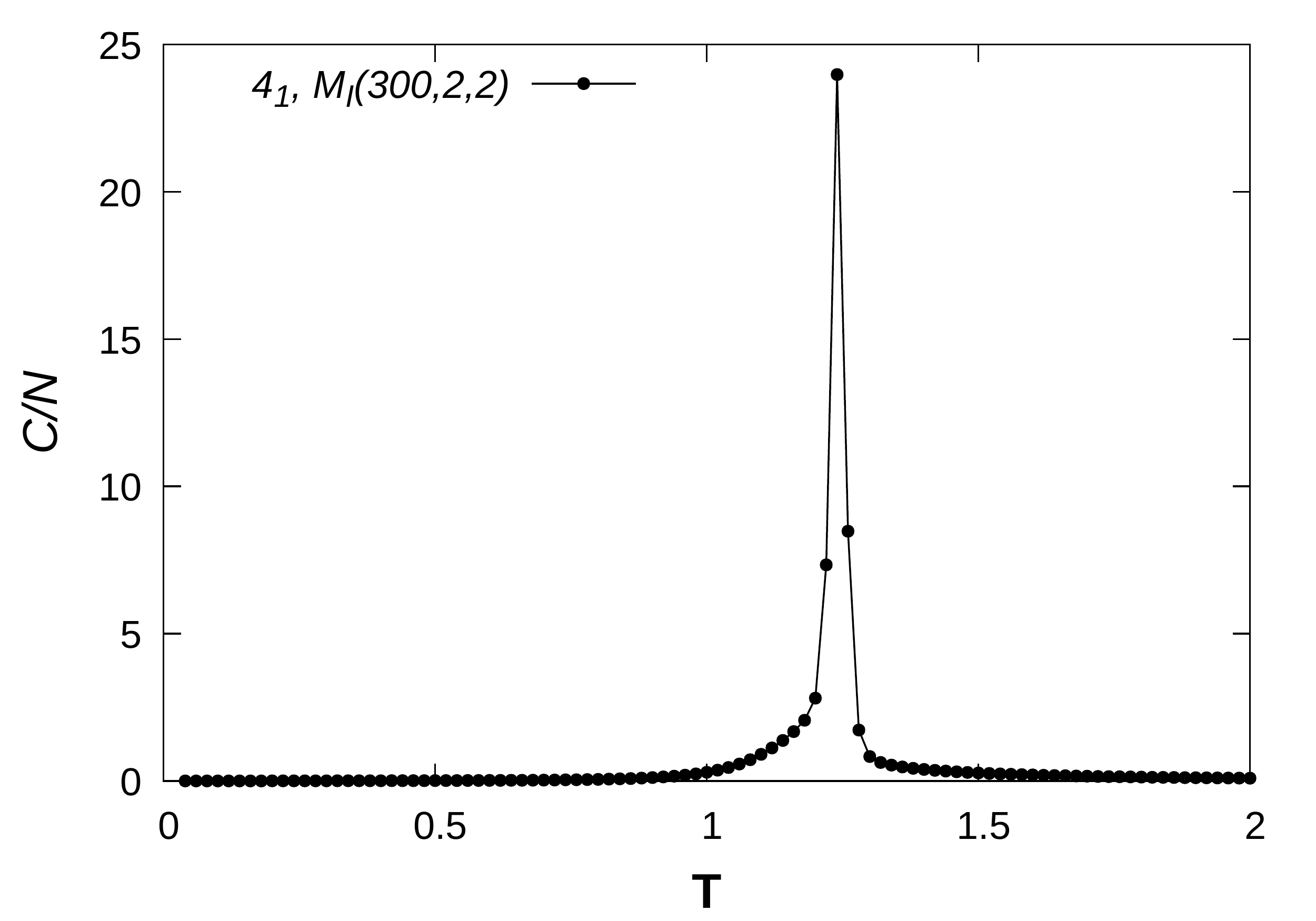}
    \caption{Left panel: Plots of the specific heat capacity of diblock copolymers
  with $N=300$ for knots: $0_1$,  $4_1$, $5_1$ and $6_2$ with monomer
  configuration $D_I(250,50)$. The plot of  knot $5_1$ with 
  configuration $D_I(200,100)$ (black squares) has been added to show the differences when the  number of monomers of type $A$  and $B$ become comparable. As we see,  the specific heat capacity of this knot exhibits just a single peak that is higher than those of the knots with
   $D_I(250,50)$. Moreover, the peak appears at a much higher
  temperature with respect to all other knots in which $N_A=250$ and $N_B=50$.
The strongest observed compact states have been observed in the case of the monomer distribution
$M_I(N,2,2)$ (right panel).}\label{fig-new-events}  
\end{center}
\end{figure}

\begin{figure}
  \begin{center}
    \includegraphics[width=0.48\textwidth]{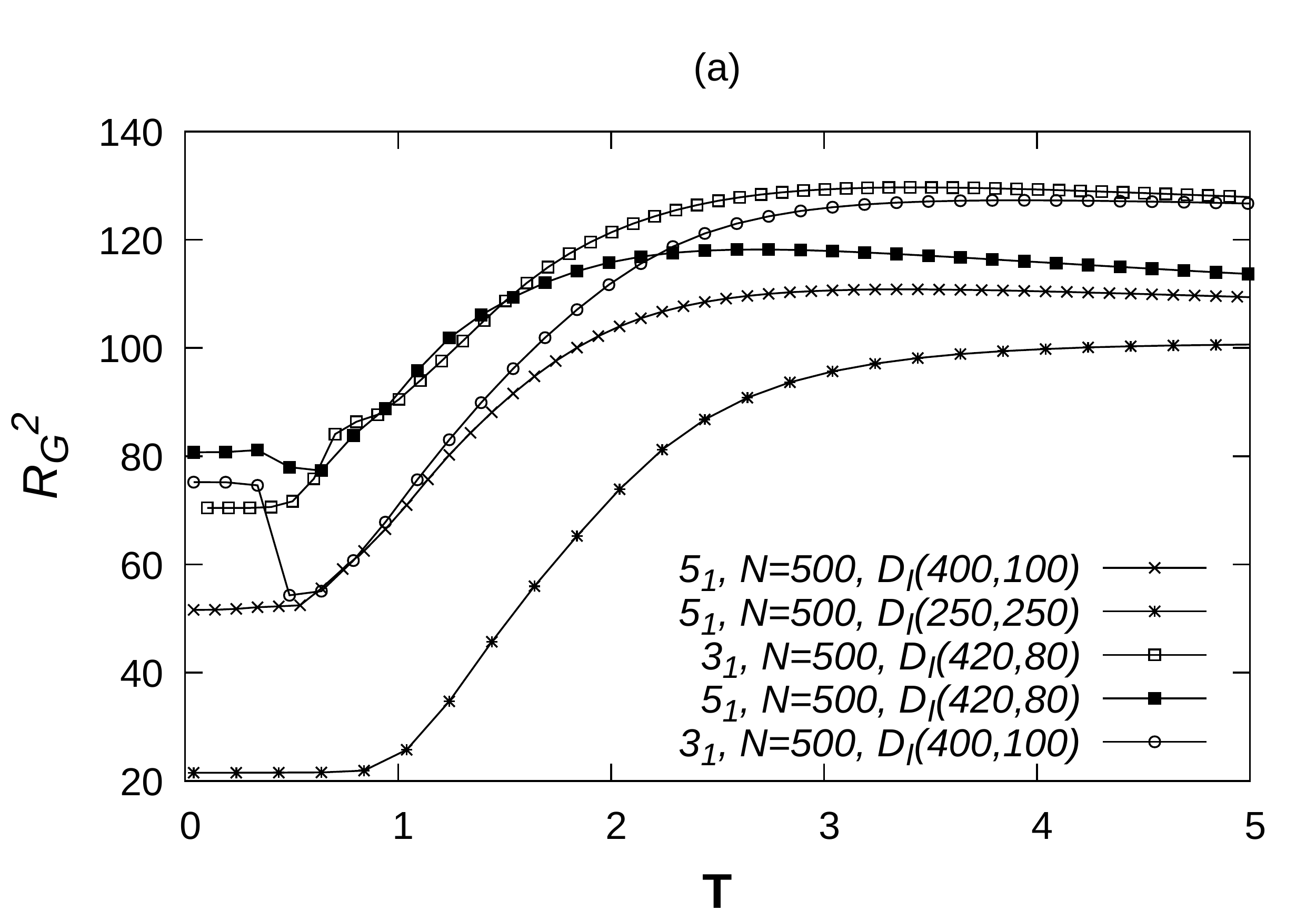}
    \includegraphics[width=0.48\textwidth]{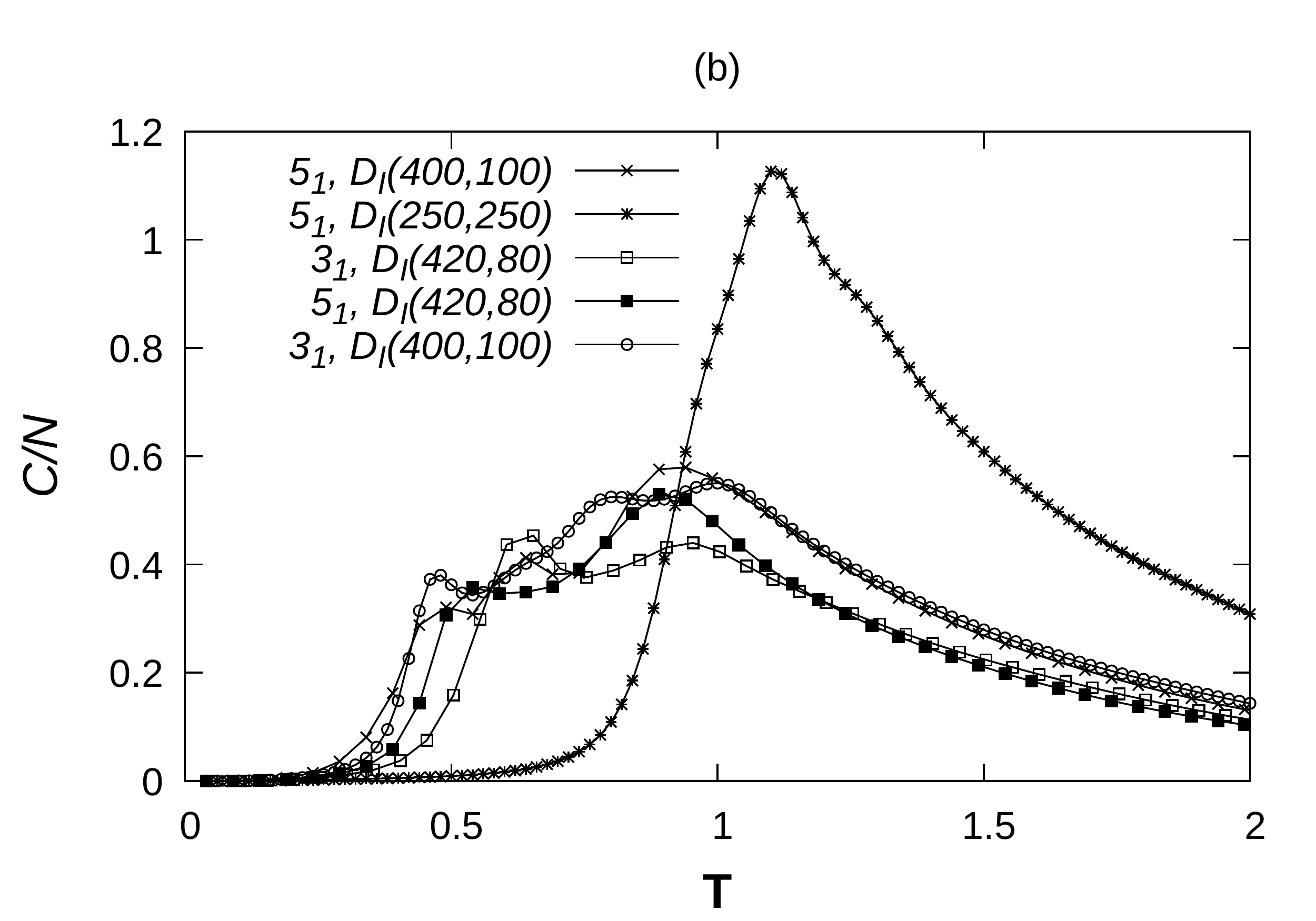} 
\caption{Plots of the gyration radius (left panel) and specific heat capacity (right panel) 
of a few knots with  length $N=500$ in three different monomer
distributions.}\label{fig-N500}  
\end{center}
\end{figure}

The data
of longer knots with $N=300$ and $N=500$ agree with the previous conclusions, see Figs.
\ref{fig-longer-shc}--\ref{fig-N500}.
In longer polymers, a third small peak appears in the
case of knot $4_1$ with $N=300$ and monomer distribution $D_I(250,50)$, see Fig.~\ref{fig-new-events}, left panel. As
explained before, this extra peak could be related to the presence of a metastable state.
In the inset of Fig.~\ref{fig-longer-shc} it is shown that
the rapid growth of the gyration radius at about $\bos T\sim 0.35$ corresponds to the first peak in the specific heat capacity.
The middle peak at about $\bos T\sim 0.6$ corresponds to a temperature in which the swelling rate is slowing down considerably, see the inset.
We interpret this with the fact that the knot is captured into a metastable state, which stabilises the size of the knot over a small interval of temperatures.
 One should keep in mind  that the detection of metastable states in long knots
 is particularly difficult.
 Indeed, there are hints that the energy landscape for long polymers
 could be funnel-like like in proteins~\cite{FFYZRMP}.
 Thus, long knots are complex systems and the search for metastable states requires an extended sampling before they are found. 
 %arrived here on 31.05.2022
% Despite the technical limitations, the density of states of the knot
% $4_1$ with 
% length $N=300$ and monomer distribution $D_{I}(250,50)$ of
% Fig.\ref{fig-longer-shc} has been derived with very high precision
% after an extended sampling.
% As it is possible to check, the number of small peaks has increased
% to three.

Finally, also the data of the gyration radius in the case of polymers
with $N=500$
confirm the general picture presented before, see Fig.~\ref{fig-N500}. Again, knots with $f\sim 0.50$ ($N_A\sim N_B$) behave differently from those with
$f\sim1$ ($N_A>>N_B$) or, equivalently, $f\sim 0$ ($N_B>>N_A$). 
As a curiosity, from the performed numerical experiments it turns out
that the  
minimal energy state created by a knot at the lowest
temperature is not always the most compact one. Indeed,
the size of the knot in two cases (knots $5_1$ 
with monomer distribution
$D_I(420,80)$ and $3_1$ with monomer distribution $D_I(400,100)$)
decreases at about $\bos T\sim 0.6$.
This is probably due to the excess of monomers of a
given type. Indeed, when temperatures are low
there are two competing conditions that should be fulfilled in order
to minimise the energy.
First, the largest possible number of
$A$ monomers should be in contact with the few  avaliable $N_B$
monomers. At the same time, however,
the $A$ monomers cannot get near to each other, as this
is energetically expensive due to the repulsive interactions between
monomers of equal type. This last requirement is responsible for the
fact that the minimal energy state could be not the most compact one.
If the temperature is rising, in fact, more energy will be available to the system,
so that conformations of the knot in which more $A$ monomers are in
contact with themselves become possible. As a consequence, 
certain knots
may attain at higher temperatures a total gyration radius which is
smaller than that of the lowest energy state.

To conclude this subsection,
we would like to stress that,
while homopolymers are  simple
systems whose size
steadily increases (in bad solvents) or decreases (in good solvents)
with growing temperatures, diblock copolymers with $f\sim 1$ (or $f\sim 0$),
exhibit a more complex behavior. Their
mean square gyration radius is smallest at low temperatures and  
increases up to its maximum value at intermediate temperatures. After
that, it starts to decrease and finally stabilizes
to some value between the maximum and the minimum at high
temperatures.
The presence of three different
regimes, compact, ultra swollen and swollen is strongly dependent on
the monomer composition. 
%When $f\sim 1$ or $f\sim 0$,
% there is an excess of monomers of type $A$ or $B$ and the repulsive
% interactions are strong.
%Yet, these knots are not behaving as homopolymers in a good solvent.
% The latter are only able to shrink slightly at high
% temperatures. Instead, copolymer knots with  $f\sim 1$ or $f\sim 0$
%expand at low temperatures like homopolymers in a bad
% solvent and shrink like homopolymers in a good solvent at higher temperatures.
%On the contrary, knots with
%$f\sim \frac 12$ have  a first expansion phase in which
%the value of their gyration radius exhibits a relevant increase. Next,
%at very high temperatures a slow expansion  is following
%until the maximum size is achieved. This behaviour is similar to that of
%homopolymers in a bad solvent. These features distinguishing knots
%with $f\sim\frac 12$ from those with $f\sim 0$ or $f\sim1$ are present 
% both in short knots with $N=90$ and in  longer knots with $N=200$,
% as it is possible to see in Figs.~\ref{fig-top-N90},
% \ref{fig-top-shc-N90}, \ref{fig-top-rg-N200} and~\ref{fig-top-shc-N200}. 

\subsection{Results for Setup II}\label{thermalsetupII}
\begin{figure}
  \begin{center}
    \includegraphics[width=0.48\textwidth]{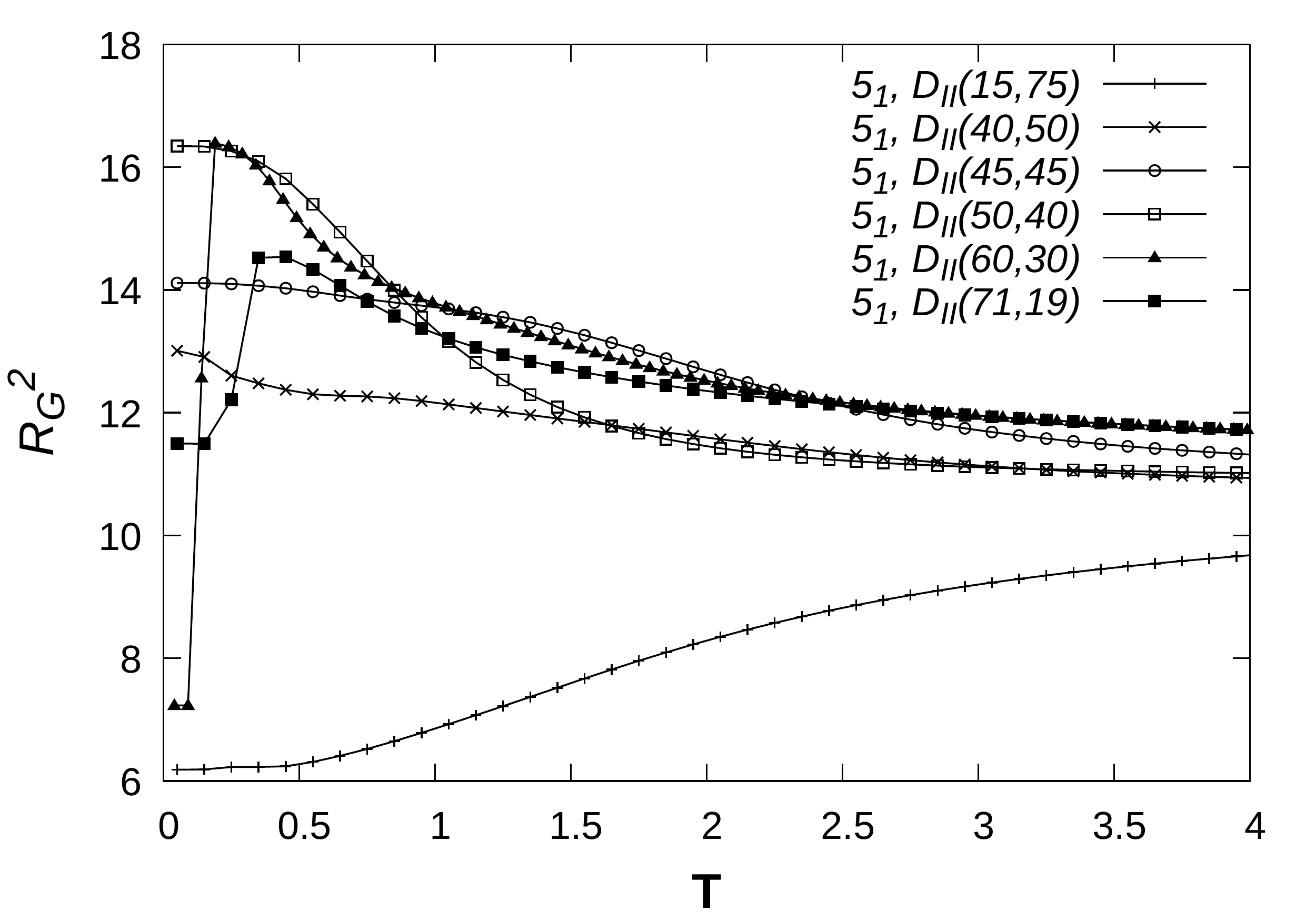}
    \includegraphics[width=0.48\textwidth]{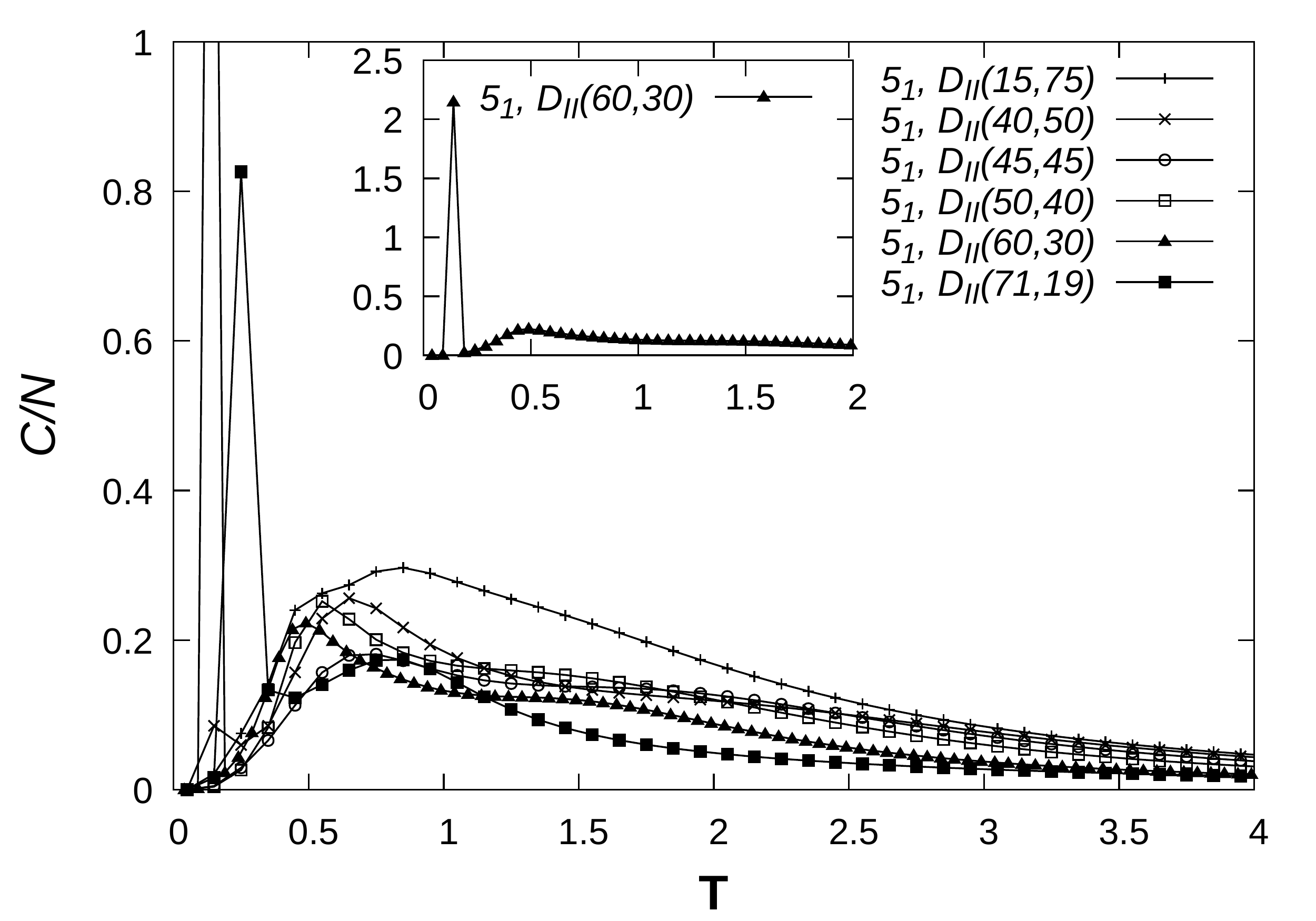}
    \caption{The mean square gyration radius $R_G^2$ (left panel) and
      the heat capacity (right panel) for a knot $5_1$ in Setup II
      with $N=90$ and different types of monomer distributions.}\label{5.1-setupII} 
\end{center}
\end{figure}

%In the case of the runs with monomer distribution 15,75 we find a
%bound state at very low temperatures formed by the 75 B monomers
%subjected to attractive interactions. That state melts at very low
%temperatures $~45 =\bos T$ and after that the A monomers, subjected
%to repulsive interactions, get nearer. This causes a small increase
%of the energy as it expected because we are at higher
%temperatures. At the temperature of about $\bos T\sim 55$ this fading
%of the repulsive interactions causes a small shrink of the entire
%knot driven by the shrinking of the A monomers only. After that, the
%expansion of the knot with growing temperatures follows up and ends
%up at very high temperatures, similarly to those of homopolymers in a
%good solvent. 
In Setup II  the monomers of type $A$ repel themselves, while monomers
of type $A$ and $B$ are subjected to excluded volume forces. Only the monomers of type $B$
attract themselves. For that reason,
it could be expected that the behavior of
knots in Setup II will be similar to that of the purely repulsive case (i. e. they attain the largest size at very low temperatures to minimize the energy and then shrink upon heating)
unless the number $N_B$ of monomers $B$ will be sufficiently high
to trigger some behavior typical of attractive interactions (i. e. they swell when heated).
This
expectation is only partially true. For instance, knots formed by diblock-copolymers 
with $N_B<<N_A$ exhibit a phase of fast, but moderate expansion when heated, so in this sense they share some properties of knots in a bad solvent despite the fact that the monomers of type $A$ subjected to repulsive forces constitute an overwhelming majority.
%the size of the knot is smallest at low temperatures despite the fact that 
%By increasing the temperature
%a moderate swelling of the knot  is observed, followed by shrinking at even higher temperatures when thermal fluctuations are dominating.
 On the contrary, when $N_A= N_B$ the size of the knot does not exhibit any significant change.
 If instead $N_B>>N_A$, attractive interactions are overwhelming and
the behavior of the knot becomes similar to that of a polymer in a bad solvent.
In all cases,
including the monomer distribution $M_{II}(N,2,2)$,
the swelling is much less marked than in Setup I. It takes place in a limited range of temperatures and it is soon followed by the shrinking which is typical
of homopolymers in a good solvent.
The situation is well summarized by
Fig.~\ref{5.1-setupII}
in which the mean square gyration radii and the specific
heat capacities of a knot $5_1$ of length $N=90$ are displayed for different monomer distributions.
A possible explanation of the different behaviours between the cases $N_A>>N_B$
and $N_A\sim N_B$
is that in the latter case the $B$ monomers form a larger number of contacts. The compact state that arises in this way is stable under temperature changes and it starts to melt only at high temperatures, i. e. when $\bos T\sim 1.00$ similarly as the powerful compact states built by homopolymers in a bad solvent. At such high temperatures the melting process and the shrinking process are no longer well distinguishable in the plots of the specific heat capacity, while this is possible when the monomer distribution is such that  $N_A>>N_B$ and the melting of the compat state takes places at much lower temperatures.
%which is  completely different from that of Setup I, is provided in the %caption of Fig.~\ref{cartoon}.

It is worth noticing that in the knot $5_1$
the swelling and shrinking phases are well recognizable only when the number of $B$ monomers is small. This is the case of the monomer distributions  $D_{II}(71,19)$ or $D_{II}(60,30)$, see left panel of Fig.~\ref{5.1-setupII}. This fact is also visible in the plots of the specific heat capacity of this knot (lines with black squares and black triangles in the right panel of Fig.~\ref{5.1-setupII} which are characterised by a double peak. The first peak can be associated to the swelling process with the melting of the compact state formed by the $B$ monomers and the second, at higher temperatures, to the shrinking process. Indeed, the first peaks in the case of the monomer distributions  $D_{II}(71,19)$ and $D_{II}(60,30)$ are centered more or less at the temperatures in which the swelling phase is taking place. The second peaks are instead much broader and appear in the range of temperatures in which shrinking takes place. Finally, the heights of the peaks are of the expected order, with higher peaks in the case of swelling and lower in the case of shrinking.

\begin{figure}
  \begin{center}
    \includegraphics[width=0.48\textwidth]{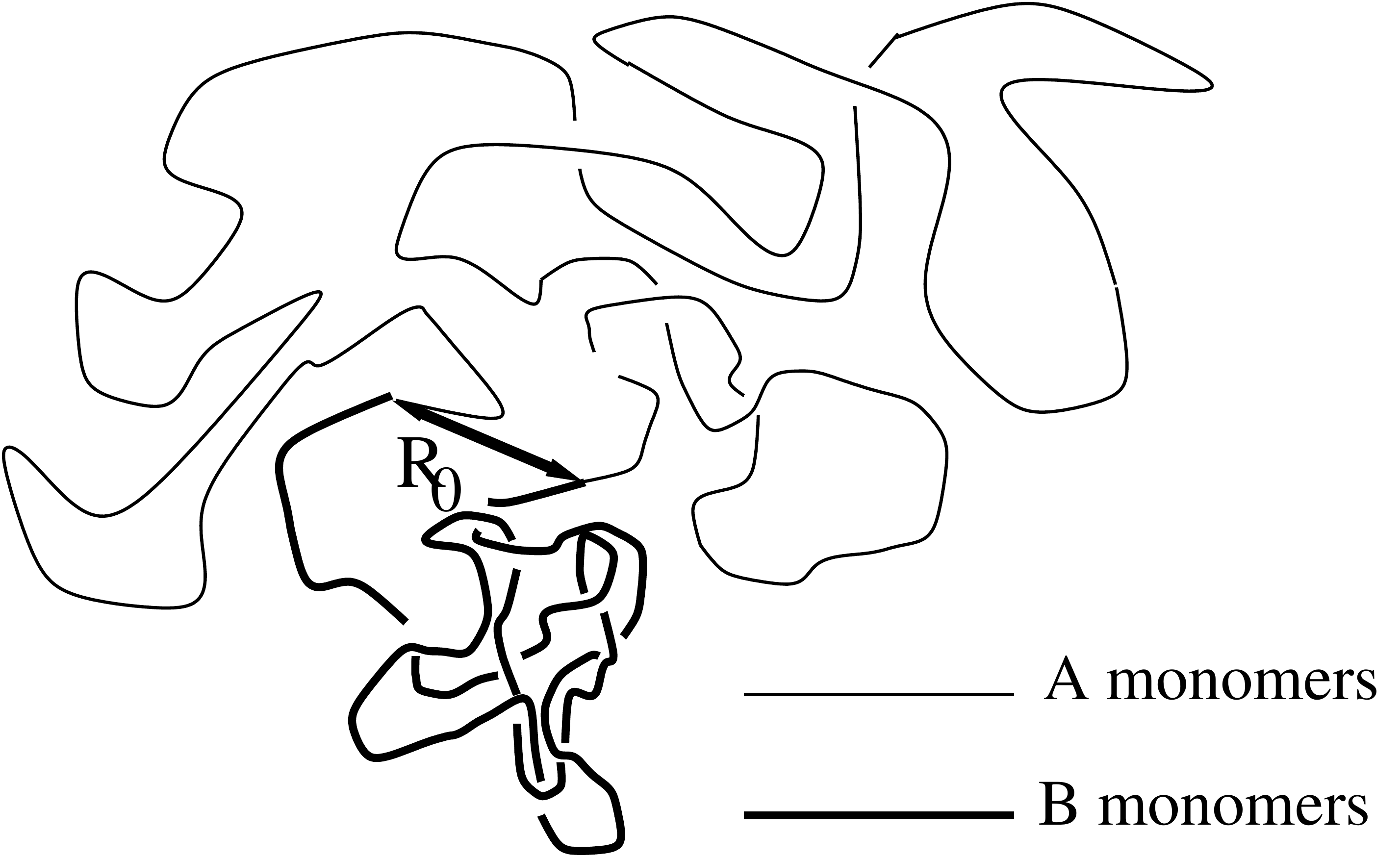}
    \caption{This figure summarises the situation of knotted diblock copolymers in Setup II: Two chains, one with $A-$type monomers and the other with $B-$type monomers, are attached together at their ends to form a knot. Both chains share the same end-to-end distance $\bos R_0$. It turns out that
      the equilibrium value of the length $|\bos R_0|$ of $\bos R_0$ is determined not only by the monomer distribution, but also by topology. The
      $A$ monomers contribute to the increase of $|\bos R_o|$ because they are subjected to repulsive interactions. On the contrary, the $B$ monomers, subjected to attractive interactions, and increasing topological complexities of the knot, both contribute to the decrease of $|\bos R_0|$. }\label{cartoon}
  \end{center}
  
\end{figure}

%The comparison of these results with the cases of pure attractive
%and repulsive  interactions can be made by looking at
%Fig.~\ref{fig-local-N90-c} in which there are the plots of the mean square gyration radii
%of two knots $3_1$ with  monomer distributions $D_{II}(0,90)$ ($f=0.00$) and
%$D_{II}(90,0)$ ($f=1.00$).

\begin{figure}
  \begin{center}
    \includegraphics[width=0.48\textwidth]{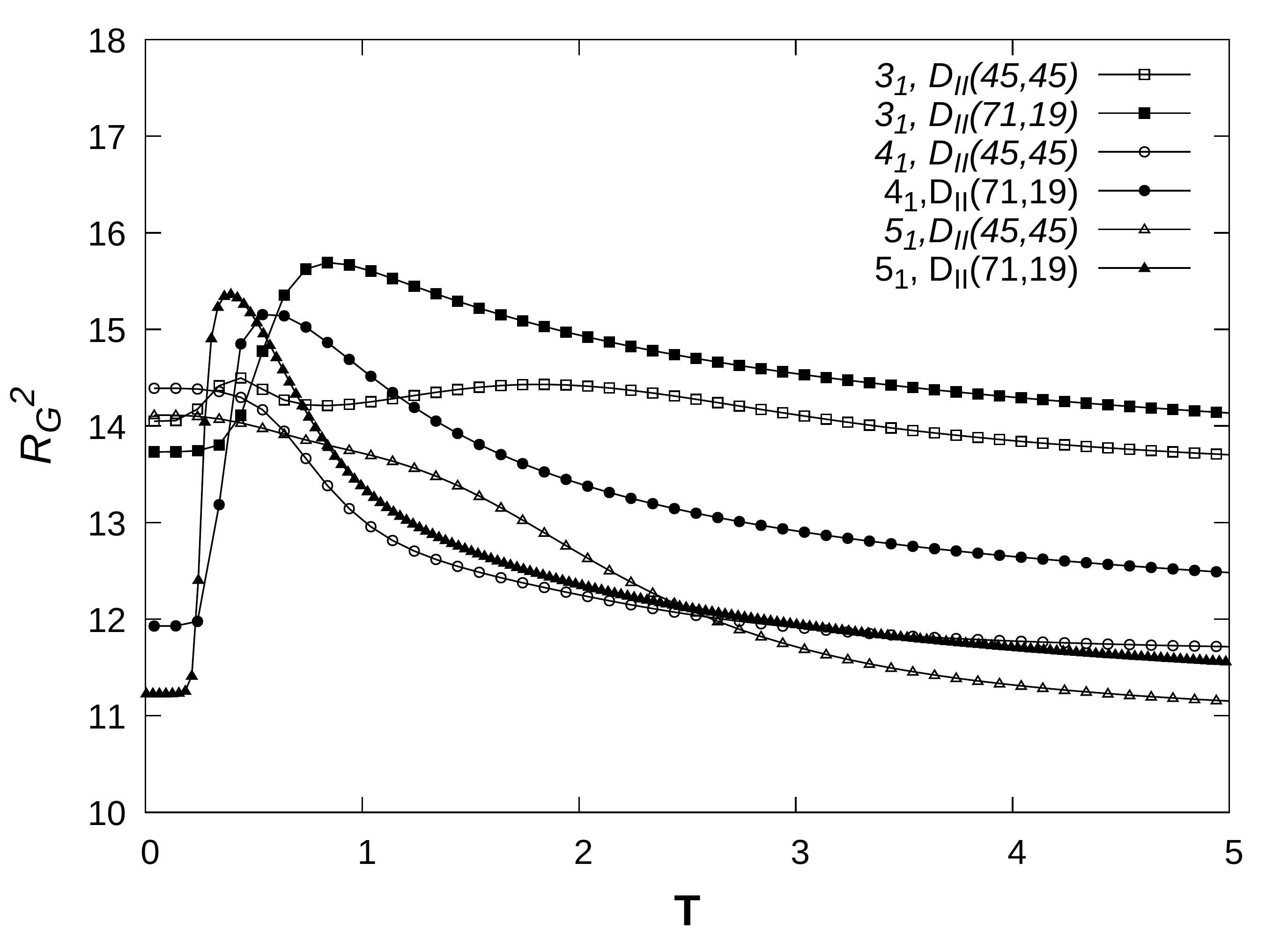}
    \caption{The mean square gyration radius $R_G^2$ of knots of
      different topology and monomer distributions in Setup II.  The length of all
      knots is  $N=90$. The plots with black points (black squares, circles and triangles) correspond to knots with monomer conformations such that $N_A>>N_B$, while those with white points (white squares, circles and triangles) correspond to knots in which $N_A\sim N_B$. In the former case, it is possible to distinguish an expansion of the knot followed by a shrinking phase.
      When $N_A\sim N_B$, only shrinking is observed or, at most, small size fluctuations (knot $3_1$, white squares). There are strong effects of topology
that may be easily detected by looking separately at the plots with black and white points. }\label{fig11} 
\end{center}
\end{figure}

\begin{figure}
  \begin{center}
    \includegraphics[width=0.50\textwidth]{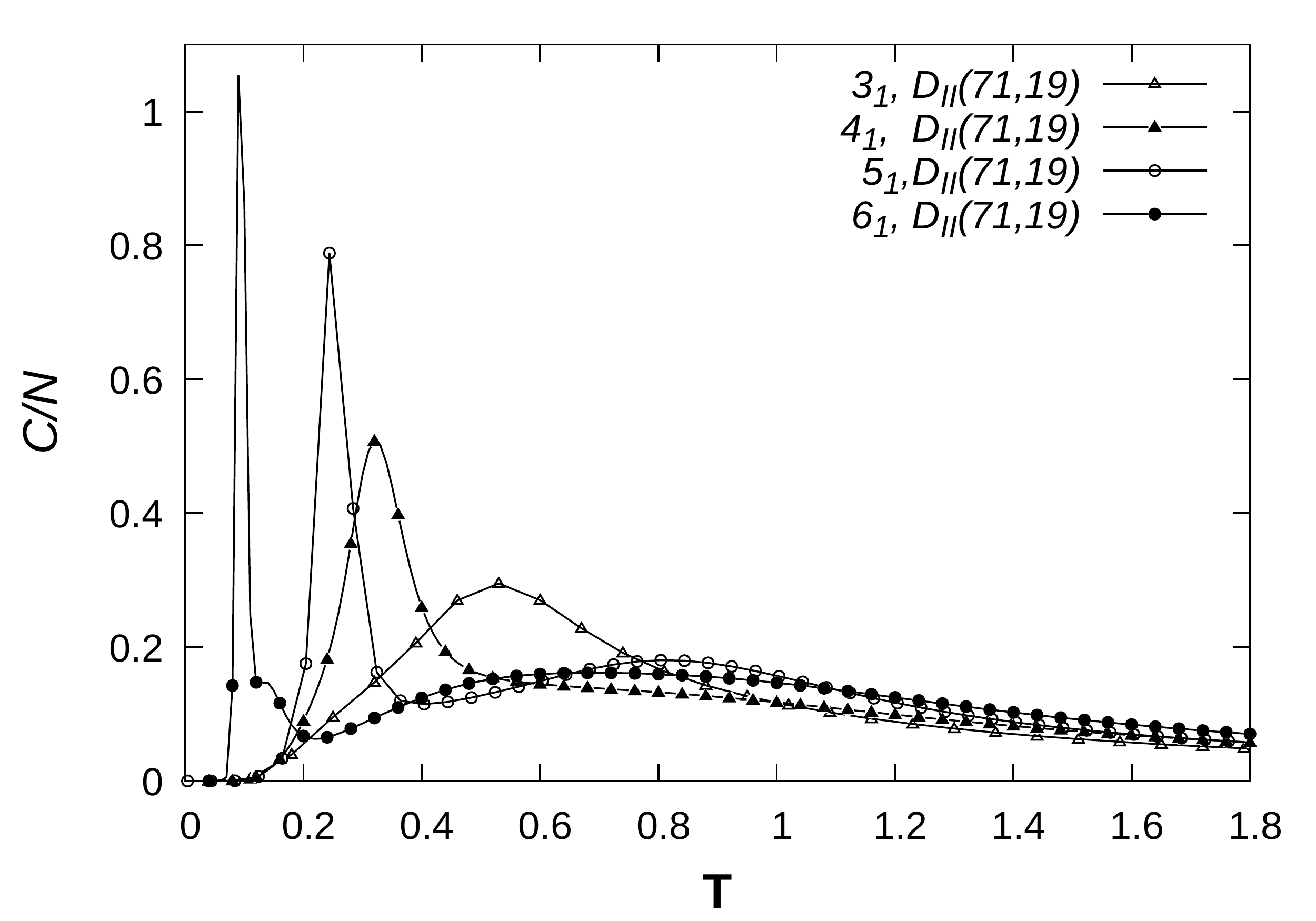}
    \caption{The specific heat capacity $C/N$ of knots  of
      different types and monomer distributions.  The length of all
      knot is  $N=90$. This picture shows that there are important effects of topology in the behaviour of the knots.}\label{fig12} 
\end{center}
\end{figure}

To check the effects of topology, in Fig.~\ref{fig11} we have displayed the gyration radii of different knots with two monomer compositions, namely
$f=0.50$ ($D_{II}(45,45)$ and $f=0.79$ ($D_{II}(71,19)$).
 As it is possible to see, also the knots $3_1$ and $4_1$ exhibit the same behaviors already observed in Fig.~\ref{5.1-setupII} in the case of knot $5_1$. However, as a general trend, it turns out that
 the swelling process taking place in knots with monomer composition  $f=0.79$ becomes increasingly more
 abrupt and start at lower temperatures with growing topological complexity. 
This influence of topology is also visible in the plots of $C/N$
of Fig.~\ref{fig12}.
As a matter of fact, when $f=0.79$,
the heights of the peaks of the specific heat capacity is gradually
rising passing from the knot $3_1$ to the knot $6_1$. Moreover, the
temperature around which the peak  of $C/N$ is centered is decreasing with increasing  knot complexity.
We notice in Fig.~\ref{fig12} that knot $4_1$ with monomer distribution $M_{II}(90,2,2)$ undergoes a swelling process that is much milder than that  of the same knot in Setup I.

%arrived here on 11.06.2022

\begin{figure}
  \begin{center}
    \includegraphics[width=0.48\textwidth]{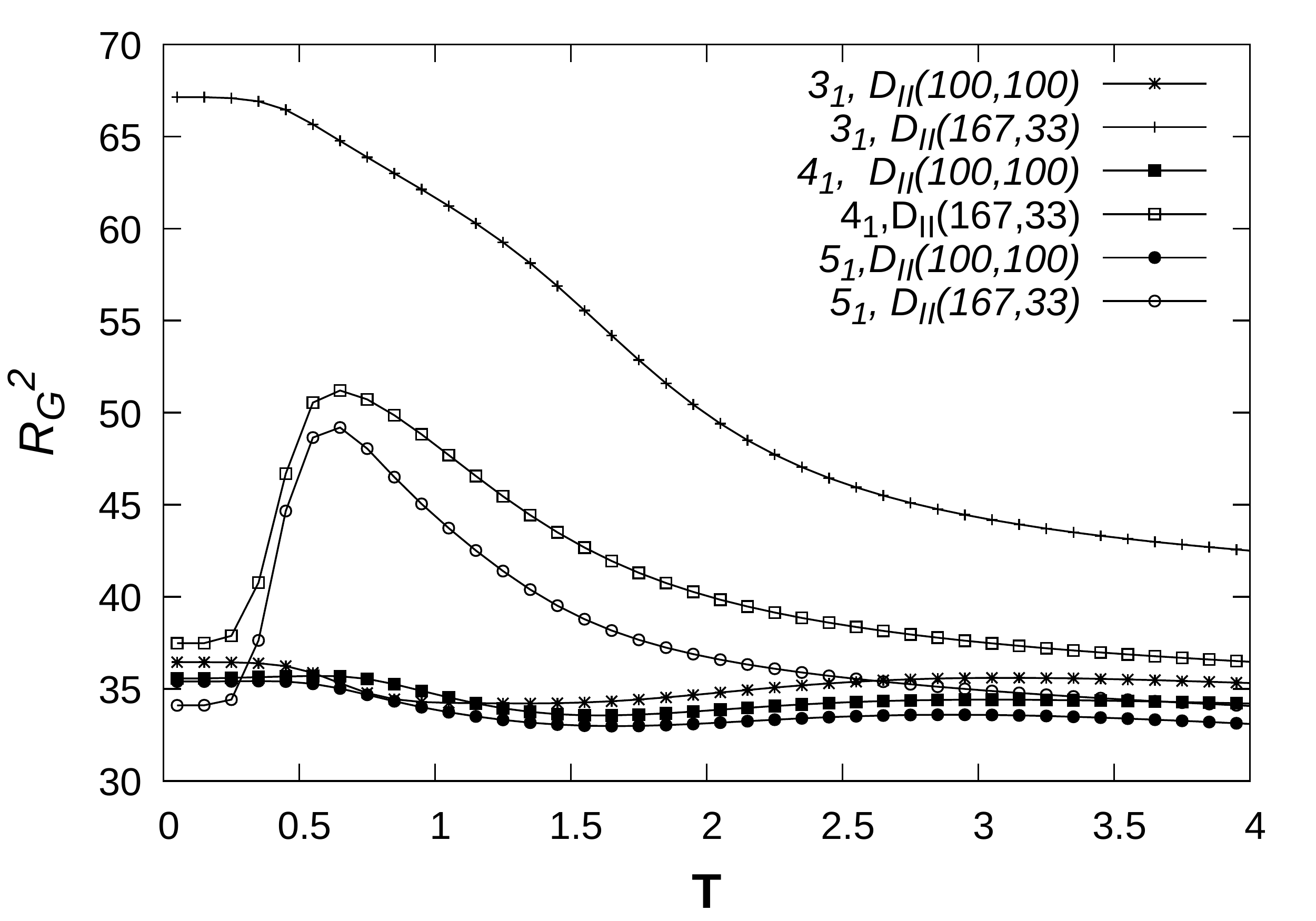}
    \includegraphics[width=0.48\textwidth]{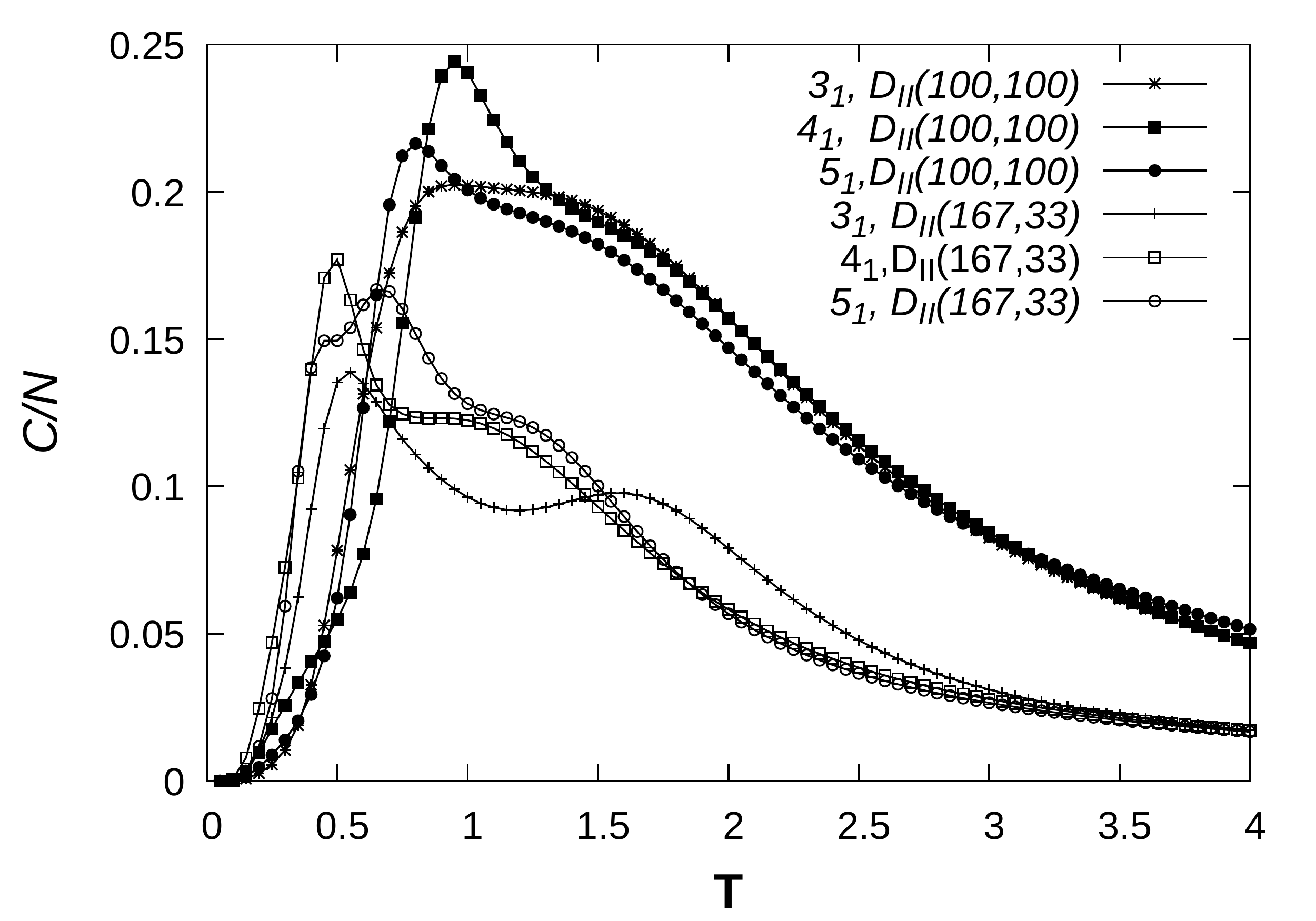}
    \caption{The mean square gyration radius $R_G^2$ of knots of
      different types and monomer distributions.  The length of all
      knots is  $N=200$.}\label{fig-setup-II-N200} 
\end{center}
\end{figure}

Going to longer knots, we see in general a fading out of the effects of topology. This is for instance visible in the fact that the knots with monomer distributions $D_{II}(100,100)$ in Fig.~\ref{fig-setup-II-N200}  have more or less the same behaviour. A remarkable exception is the knot $3_1$ with monomer distribution $D_{II}(167,33)$ whose values of the gyration radius are decidedly greater than those of knots $4_1$ and $5_1$ with the same monomer distribution. This effect is certainly due to topology
and it has been observed also in the case of the knot $3_1$ with length $N=90$ and monomer distribution $D_{II}(60,30)$.
A temptative explanation of this phenomenon can be the following. Looking at the picture in Fig.~\ref{cartoon}, we see that in Setup II knots consist into two open chains, one with monomers of type $A$ and one with monomers of type $B$, joined together at their ends. For this reason, the end-to-end distance $R_0$ is common for both chains.
$R_0$ determines to some extent also the gyration radii of these chains and eventually the gyration radius of the whole knot.
Clearly, the segment with the $A$ monomers, which are subjected to repulsive interactions, will try to increase the value of $R_0$. On the contrary, the segment with the $B$ monomers which are attracting themselves, will tend to have smaller values of its end-to-end distance. If the monomer distribution is $D_{II}(167,33)$, then the repulsive interactions will certainly be dominating because of the large number of $A$ monomers. This would imply that the value of $R_0$ will be mainly determined by the part with the $A$ monomers. However, if the topology of the knot is complex, then the numbers of turns made by the path of the segment with the $A$ monomers will be high. This will make the knot more compact and thus also $R_0$ will be relatively smaller than in simpler topologies. In this situation it is very likely that the effects of the fluctuations to which the $A$ monomers are subjected will be hampered by the topological constraints and will not be able to destroy the contacts made by the $B$ monomers.
If this happens, there is a chance that the the $B$ monomers will prevail and succeed to keep the value of $R_0$ small as required by the energy and entropy considerations for segment $B$. 
As far as it is possible to see from our simulations, this topological mechanism to keep together the knot in a compact state is
working when the topology is more complex than that of a knot $3_1$.

%\begin{figure}[h]
%\begin{center}
%\includegraphics[width=0.5\textwidth]{fig-convergence-rg.pdf}
%\caption{The gyration radius $R_G^2$ of the knot $7_1$ with $N=90$  and
%  copolymer type $T_1(45,45)$ is plotted for
%  increasing widths of the considered energy interval $\cal
%  I$. } \label{fig-convergence-tris}   
%\end{center} 
%\end{figure}
\section{Conclusions}\label{conclusions}
The Wang-Landau algorithm has been applied here to study the thermal properties of knotted copolymers in a solution.
Two different types of monomers have been considered, called type $A$ and type $B$. The monomers are  subjected to very short-range interactions. Two different setups have been investigated. Setup~I corresponds to the case of charged  monomers in an ion solution. The $A-$type monomers carry a positive charge and $B-$type monomers a negative one. In Setup~II the monomers are not charged, but the solvent is good for type $A$ monomers and bad for the $B$ monomers.
Block copolymers knots exhibit a more complex behaviour than knotted homopolymers. The latter are in fact simple
two-state systems. When they are in a bad solvent, they are found at
very low temperatures in extremely compact and ordered conformations,
called crystallites following \cite{binder}. With the increasing of
the temperature, these knots start to swell until the state of
expanded coils is reached. The swelling process is much less violent
than in the case of linear polymer chains studied in \cite{binder},
whose specific heat capacity is characterized by a very sharp peak. In
knotted polymer rings, the peak caused by swelling is much broader.
When the solvent is good,
instead, 
knots made by homopolymers are in an expanded  state
at low temperatures and slightly shrink with increasing temperatures, see \cite{yzff,yzff2013}.

The introduction of monomers of two kinds drastically changes this situation.  
For example, for suitable values of the monomer composition $f$,
knots formed by $AB-$diblock copolymers in an ion solution (Setup I)
perform as homopolymers in a bad solvent at low temperatures, but
exhibit features typical of homopolymers in a good solvent at high
temperatures.  
In practice, they become systems with several states in which at least three different states can be distinguished.
At low temperatures these knots are found
in the compact and ordered state of crystallites. With growing temperatures, they are swelling
like homopolymers in a bad solvent, but  after a maximum gyration
radius is reached, they start to shrink like knotted homopolymer rings
in a good solvent. 
Examples of this behaviour are in Setup I the knot $3_1$ with monomer distribution $D_I(75,15)$ of Fig.~\ref{fig1a} and in Setup II the knot $5_1$ with monomer distribution $D_{II}(60,30)$ of Fig.~\ref{5.1-setupII}.
By choosing the topology and the monomer composition of the knot, it is
possible to tune both its size  at different
temperatures and the temperature at which the maximum value of the
gyration radius is attained. Also the range in which the gyration
radius is allowed to vary can be determined to some extent. 
Knots with such features are clearly an advantage with respect to homopolymers
in  potential medical applications and in the production of intelligent
polymer materials containing knots.

In Setup II, knots formed by $AB-$diblock
copolymers  have a behavior that is strongly dependent on the
monomer composition $f$. Following the intuition,
if the number of $A$ monomers largely exceeds that of $B$ monomers, i.~e. $N_A>>N_B$, then it could be expected that
knots behave like knotted homopolymer rings in a good solvent, since
the $A$ monomers are subjected to repulsive interactions.
Conversely, if $N_B>>N_A$ and the $B$ monomers are below the theta point, we would rather expect a behavior typical of a homopolymer knot in  a bad 
solvent. The performed simulations show however that the situation is more complicated than that.
For instance, in  the case $N_A>>N_B$ we see from Fig.~\ref{fig11} that,
in the case of the knots $3_1,4_1$ and $5_1$ with monomer distribution $D_{II}(71,19)$,
the less numerous $B$ monomers play the dominant  role at very low temperature. Indeed, these knots
exhibit features typical of homopolymers in a bad solvent, i.~e. they swell
with growing temperatures.

Multiblock copolymers in the classes $M_I(N,n_A,n_B)$ and $M_{II}(N,n_A,n_B)$, where $n_A$ and $n_B$ are small in comparison to $N$, have remarkable
properties too. These properties are not easily predictable by simply looking at
the polymer composition. 
For example, we observe a
transition from the compact state to the swollen state which is much more abrupt
than that of knots realized using monomers of the same type or diblock
copolymers.

The main conclusions of this work can be summarised as follows:
\begin{enumerate}
%highlights:
\item The strongest compact states due to the 
  contacts formed by the monomers subjected to attractive interactions have been observed in knots of various lengths with 
  monomer distribution $M_{I}(N,2,2)$, see Figs.~\ref{fig-local-N90-b}, right panel and \ref{fig-new-events}, right panel. The specific heat capacity
  of these knots is characterised by a high peak  concentrated in a very narrow range of temperatures corresponding to the
melting of these bound states. This interpretation is corroborated by the fact that, exactly in the same range of temperatures, the knot undergoes a
sudden and rapid swelling process.
When monomer distributions of the kind $M_{I}(N,n_A,n_B)$ are considered with increasing  values of $n_A$ and $n_B$
($n_A,n_B=4,8,...$), the peak widens and its height becomes lower, implying that
strong compact states are still present at least up to the tested value of $n_A=n_B=8$, but they become weaker and weaker.
\item  Topological effects strongly influence the behaviour of knots in the case of short polymers ($N\sim 90$). Examples of these effects in Setup I can be observed in
the plots of the gyration radius
and specific heat capacity reported
in Fig.~\ref{fig-top-N90} and in Fig.~\ref{fig-top-shc-N90} respectively. Fig.~\ref{fig12} displays the changes due to topology of the heights and temperatures of the peaks of the specific heat capacity in the case of Setup II.
With increasing polymer lengths these effects fade out and
the choice of the monomer distribution becomes the main factor influencing
the properties of the knots. Yet, topology is still relevant for longer polymers because it provides a way to fine tune their behaviour.
For instance, the effects of topology when $N=200$ on the way in which the gyration radius
changes with different temperatures are reported in Fig.~\ref{fig-top-rg-N200}. A certain dependence on topology of
the heights of the peaks of the specific heat capacity
can also be spotted in the plots of Fig.~\ref{fig-top-shc-N200} where
knots of length $N=200$ are considered.
However, in general the influence of topology is more visible in the plots of the gyration radius. Let us notice that topology plays some role even in the case of the longest polymers that have been investigated here.
For instance, the knot $3_1$ with $N=500$ and monomer distribution $D_I(400,100)$ is suddenly shrinking at a temperature of $\bos T\sim 0.5$ (the value of $R_G^2$ goes from $75$ to $52$), a feature that is not so marked in the case of the other knots whose plots have been displayed in Fig.~\ref{fig-N500}, left panel, including the knot $5_1$ with the same monomer distribution.
\item  Several exceptions to the rule that the influence of topology should fade out with increasing polymer lengths have been observed.
  For instance, the gyration radius of the knot $3_1$ with length $N=200$ and monomer distribution $D_{II}(167,33)$ is much larger than
  the gyration radii of knots $4_1$ and $5_1$ with the same lengths and monomer distribution, see Fig.~\ref{fig-setup-II-N200}. This striking difference can be explained by the strong entropic effects induced in Setup II
by the fact that  the paths of the knots are subjected to topological constraints, see Fig~\ref{cartoon} and related comments. Also when $N=90$, it is possible to see from Fig.~\ref{fig11} that, when the monomer distribution is $D_{II}(71,19)$, the swelling of the knot $3_1$ with rising temperatures is much more limited than that of knots $4_1$ and $5_1$.
\item A characteristics that emerges in knotted block copolymers
and is not present in the case of homopolymers,
is the existence of rearrangements of the knot structures at low temperature. In some cases, this leads to intermediate states. These rearrangements can be detected by the appearance of extra peaks or shoulders in the specific heat capacity of longer knots in Setup~I.
Metastable intermediate states have been observed  for instance in knot $3_1$ with $N=500$ and monomer distribution $D_I(400,100)$, see Fig.~\ref{fig-N500}, right panel, and knot $4_1$ with $N=300$ and monomer distribution $D_I(250,50)$, see Fig.~\ref{fig-new-events}.
\end{enumerate}
%finish these conclusions describing the algorithm used in order to tackle the calculations, how the numbers of contacts help in determining the conformations of the knots etc.
  The
simulations presented in this paper require the sampling of an
extensive amount of knot conformations. Despite major improvements
in the sampling procedure, that of rare events is still a problem in the case of very long polymers.
Some of the 
conformations
appear after several hundred billions of trials and their inclusion 
extends enormously the calculation times.
Moreover, in this work very short-range interactions have been considered. This is enough to study the cases of flexible knots in a good or bad solutions, but it would be interesting to add more complicated interactions.
In this way it would be possible to consider for instance also the polymer rigidity and the transition from bad to good solvents at the theta point.
Work is in progress to implement in our code the backbone rigidity and the Lennard-Jones interactions.
\begin{acknowledgments} 
The simulations reported in this work were performed in part using the HPC
cluster HAL9000 of
the University of Szczecin.
The research presented here has been supported by the Polish National Science Centre under
grant no. 2020/37/B/ST3/01471.
This work results within the collaboration of the COST
Action CA17139 (EUTOPIA). The use of some of the facilities of the Laboratory of
Polymer Physics of the University of Szczecin, financed by 
a grant of the European Regional Development Fund in the frame of the
project eLBRUS (contract no. WND-RPZP.01.02.02-32-002/10), is
gratefully acknowledged.  \end{acknowledgments}
%\section{Bibliography}
 
\end{document}